\documentclass{article}

\usepackage{microtype}
\usepackage{graphicx}
\usepackage{subcaption}
\usepackage{booktabs} 

\usepackage{hyperref}

\usepackage[main, final]{neurips_2026}

\usepackage{amsmath}
\usepackage{amssymb}
\usepackage{mathtools}
\usepackage{amsthm}
\usepackage{graphicx}
\usepackage{makecell}
\usepackage{subcaption}
\usepackage{amsmath} 
\usepackage{comment}
\usepackage{float}
\usepackage{bbding}
\usepackage{tablefootnote}
\usepackage[table,dvipsnames]{xcolor}
\usepackage{colortbl}
\newcommand{\ccc}[1]{\cellcolor{gray!30}#1}

\title{Rethinking Training Targets, Architectures and Data Quality for Universal Speech Enhancement}

\author{
Szu-Wei Fu \\
NVIDIA \\
\texttt{szuweif@nvidia.com} \\
\And
Rong Chao \\
Academia Sinica\\
\texttt{roychao19477@gmail.com} \\
\And
Xuesong Yang \\
NVIDIA \\
\texttt{xueyang@nvidia.com} \\
\And
Sung-Feng Huang \\
NVIDIA \\
\texttt{sungfengh@nvidia.com} \\
\And
Ryandhimas E. Zezario \\
Academia Sinica\\
{\small\texttt{ryandhimas@saturn.yzu.edu.tw}} \\
\And
Rauf Nasretdinov \\
NVIDIA \\
\texttt{rnasretdinov@nvidia.com} \\
\And
Ante Juki\'c \\
NVIDIA \\
\texttt{ajukic@nvidia.com} \\
\And
Yu Tsao \\
Academia Sinica\\
\texttt{yu.tsao@citi.sinica.edu.tw} \\
\And
Yu-Chiang Frank Wang \\
NVIDIA \\
\texttt{frankwang@nvidia.com} \\
\\
}

\begin{document}

\maketitle

\begin{abstract}
Universal Speech Enhancement (USE) aims to restore speech quality under diverse degradation conditions while preserving signal fidelity. Despite recent progress, key challenges in training target selection, the distortion--perception tradeoff, and data curation remain unresolved. In this work, we systematically address these three overlooked problems. First, we revisit the conventional practice of using early-reflected speech as the dereverberation target and show that it can degrade perceptual quality and downstream ASR performance. We instead demonstrate that time-shifted anechoic clean speech provides a superior learning target. Second, guided by the distortion--perception tradeoff theory, we propose a simple two-stage framework that achieves minimal distortion under a given level of perceptual quality. Third, we analyze the trade-off between training data scale and quality for USE, revealing that training on large uncurated corpora imposes a performance ceiling, as models struggle to remove subtle artifacts. Our method achieves state-of-the-art performance on the URGENT 2025 non-blind test set and exhibits strong language-agnostic generalization, making it effective for improving TTS training data. Model weights are available for download at: \url{https://huggingface.co/nvidia/RE-USE}. 

\end{abstract}

\section{Introduction}
Universal speech enhancement (USE)\cite{serra2022universal} aims to improve the intelligibility and perceptual quality of degraded speech across diverse conditions while preserving attributes such as speaker identity, emotion and accent~\citep{babaev2024finally}. 
Recent research has increasingly shifted from task-specific methods toward generalized models capable of handling heterogeneous domains and conditions.
For instance, VoiceFixer~\citep{liu2021voicefixer} combines a ResUNet-based analysis stage with a neural vocoder-based synthesis stage, while MaskSR~\citep{li2024masksr} employs a masked generative modeling objective to handle comparable conditions. More recently, AnyEnhance~\citep{zhang2025anyenhance} introduced prompt-guidance and a self-critic mechanism, offering a unified framework capable of mitigating diverse degradations across speech and singing voice. 
To further advance this generalization, the URGENT 2025 Challenge~\citep{saijo2025interspeech} established a standardized training data from diverse sources with varying quality covering seven distortion types (additive noise, reverberation, clipping, bandwidth limitation, codec artifacts, packet loss, and wind noise) across multiple sampling rates (8, 16, 22.05, 24, 32, 44.1, and 48 kHz) and five languages (English, German, French, Spanish, and Chinese). Adopting this rigorous setup, we investigate three previously overlooked bottlenecks: (1) training target selection, (2) the trade-off between fidelity and perceptual quality, and (3) training data curation.

\textbf{\underline{Training Targets}}: In the URGENT Challenge, six of the seven speech distortions use original anechoic clean speech as the training target. However, for \textbf{dereverberation}, the standard target is \textit{early-reflected speech}, derived by convolving original anechoic clean speech with the early reflection component of the room impulse response (RIR). 
This convention stems from the common belief that "removing early reflections is challenging and may introduce excessive artifacts" ~\citep{valin2022dereverb, zhou2023speech, zhao2020monaural}.
Contrary to this view, we found that retaining early reflections for training degrades both perceptual quality and machine intelligibility measured by downstream ASR performance. 
We argue that the primary difficulty in dereverberation is not the early reflections themselves, but the misalignment between the reverberant input and clean target caused by the implicit estimation of the \textbf{direct-path time shift}.
We propose using time-shifted anechoic clean speech as the learning target, which resolved this misalignment and significantly boosts performance. 
While previous studies have used similar targets~\citep{zhao2020monaural, wang2021convolutive}, we provide the first systematic, large-scale evaluation of these two targets under diverse degradation conditions and demonstrate the superiority of the time-shifted anechoic clean target.

\textbf{\underline{Model Architecture}}: Existing approaches often struggle to balance signal fidelity and perceptual quality. For example, in the URGENT Challenge, \citep{sun2025scaling} proposes a regression model that introduces a channel-mixing module to bridge time- and frequency-domain modeling, together with progressive block extension to enable model training at different scales. Although effective overall, its quality is inherently limited, as it tends to produce over-smoothed outputs under severe bandwidth limitations and packet loss ~\citep{saijo2025interspeech}. 
In contrast, purely generative approaches may improve perceptual quality but risk introducing hallucinated contents ~\citep{saijo2025interspeech}.
In this Challenge, hybrid methods therefore explore ways to integrate the outputs of regression and generative models.
\citep{chao2025universal} combines the two outputs using a simple energy-based criterion computed from the noisy input. Both \citep{rong2025ts} and \citep{goswami2025fuse} propose three-stage frameworks to generate the final USE output: \citep{rong2025ts} uses filling, separation, and restoration modules, while \citep{goswami2025fuse} uses a fusion network that combines regression outputs with a token sampling-based generative model. Going further, \citep{le2025multistage} applies a four-stage strategy composed of audio declipping, packet loss compensation, audio separation, and spectral inpainting. \textbf{An important question, therefore, is which combination strategy is optimal and theoretically justified}.

Motivated by the theoretical distortion-perception tradeoff (see Section~\ref{combine} and~\ref{Ap1} in the Appendix), we propose a streamlined two-stage framework that effectively combines the strengths of both paradigms. We first train a regression model to convergence (ensuring high fidelity) and freeze it. Its output then serves as a conditional input to a generative model (restoring perceptual details). This approach eliminates the complex heuristics of prior multi-stage methods while achieving optimal signal fidelity under a perceptual-quality constraint, with theoretical support provided in Section~\ref{Ap1} of the Appendix.

\textbf{\underline{Training Data Quality}}: 
Deep learning models are fundamentally constrained by their data. While scaling laws regarding data quantity are well-studied~\citep{zhang2024beyond,gonzalez2024effect}, the impact of data \textit{quality} in large-scale USE remains underexplored, with only one concurrent work hinting at its importance ~\citep{li2025less}. 
We analyze the URGENT Challenge data and find that despite organizer filtering, many ``clean'' samples contain significant residual degradations, not only due to the use of early-reflected speech (Section~\ref{Trade-off}). We demonstrate that training on such imperfect data imposes a hard performance ceiling, preventing models from removing subtle artifacts like electrical microphone hiss. We provide a detailed example of how data curation directly impacts performance to unseen real-world conditions (Section~\ref{unseen}).

In summary, our work makes the following contributions:

1. We critically reassess dereverberation learning targets, showing that time-shifted anechoic clean speech consistently outperforms early-reflection targets in USE settings.

2. We propose a theoretically grounded two-stage framework that achieves an optimal fidelity--perception trade-off.

3. We conduct a comprehensive analysis of the trade-off between training data quality and quantity, highlighting the importance of data curation.



\section{Proposed Method}
\label{gen_inst}
In this section, we analyze key challenges in universal speech enhancement and present our proposed solutions. 
First, we address the limitations of conventional dereverberation targets and justify using time-shifted anechoic clean speech.
Second, we introduce a two-stage framework that effectively combines regression and generative models to resovle the fidelity-quality dilemma.
Finally, we investigate the critical trade-off between training data scale and quality, a relatively new topic that has been seldom explored.

\subsection{Shifted Anechoic Clean Speech as a Superior Learning Target}

\label{anechoic clean}

Given an anechoic clean speech $s$ and an RIR $r$, the reverberant speech $y$ is modeled as the convolution between them:
\begin{equation}
\label{eq1}
\begin{aligned}
y[n] &= s[n] \ast r[n] 
\end{aligned}
\end{equation}
where $n$ denotes the discrete time index, $\ast$ denotes the convolution operator. The RIR $r[n]$ can be decomposed into the direct path $\delta[n-n_0]$, early reflection $r_e[n]$, and late reflection $r_l[n]$ (see Figure~\ref{fig:RIR} in Appendix):
\begin{equation}
\label{eq2}
\begin{aligned}
r[n] &= \delta[n-n_0] \ast \left( r_e[n]+r_l[n] \right)  
\end{aligned}
\end{equation}
Here, $\delta[n]$ is the Dirac delta function, and $n_0$ represents the direct-path time shift (typically 5--30 ms). We estimate $n_0$ based on the maximum magnitude of the RIR, i.e., $n_0 = \arg\max_n |r[n]|$. Early reflections are defined as RIR components occurring within 50 ms after the direct-path peak $\delta[n-n_0]$ as specified in the URGENT Challenge
~\citep{saijo2025interspeech, zhang2024urgent} and other studies~\citep{wang2021convolutive}, while subsequent impulses components constitute late reflections~\citep{naylor2010speech}.

Conventional dereverberation approaches, including the URGENT Challenge, typically use early-reflected speech $s_e[n] = s[n] \ast \delta[n-n_0] \ast r_e[n]$ as the learning target. This convention stems from the belief that \textbf{``early reflections are much harder to remove, and the difficulty of solving the problem leads to excessive artifacts in the enhanced speech''}~\citep{valin2022dereverb,zhou2023speech,zhao2020monaural}. Indeed, we found that using the unshifted anechoic clean signal $s[n]$ directly as a target yields the poorest performance~(Figure~\ref{fig:UTMos_VqScore}\mbox{(b)}), aligning with prior findings.

Although early reflections have a smaller impact on perceived quality compared to late reverberation~\citep{valin2022dereverb}, we find that retaining early reflections still significantly degrades speech quality metrics~(UTMOS~\citep{saeki2022utmos}, DNSMOS~\citep{reddy2022dnsmos}, NISQA~\citep{mittag2021nisqa}). 
As shown in Figure~\ref{fig:UTMos_VqScore}\mbox{(b)}, progressively reducing the early reflection window from 50 ms to 0 ms consistently improves quality scores of the enhanced speech. 

Setting the window to 0 ms reduces the target to the time-shifted anechoic signal of $s$, as $s[n] \ast \delta[n-n_0] = s[n-n_0]$. Consequently, in Equation~\eqref{eq2}, eliminating early reflections poses little difficulty for the model. The critical challenge is the implicit estimation of the \textbf{direct-path time shift}, which introduces a misalignment of $n_0$ between the reverberant input and the clean target. 
By using the time-shifted target $s[n-n_0]$, we effectively bypass this alignment issue. This confirms that $s[n-n_0]$ outperforms the conventional early-reflected target ($s_e[n] = s[n] \ast \delta[n-n_0] \ast r_e[n] = s[n-n_0] \ast r_e[n]$), and outperforms further the original clean speech $s[n]$. 

\subsection{Bridging Fidelity and Quality: A Two-Stage Framework}
\label{combine}

\begin{figure}[ht]
\begin{center}
\includegraphics[width=0.7\linewidth]{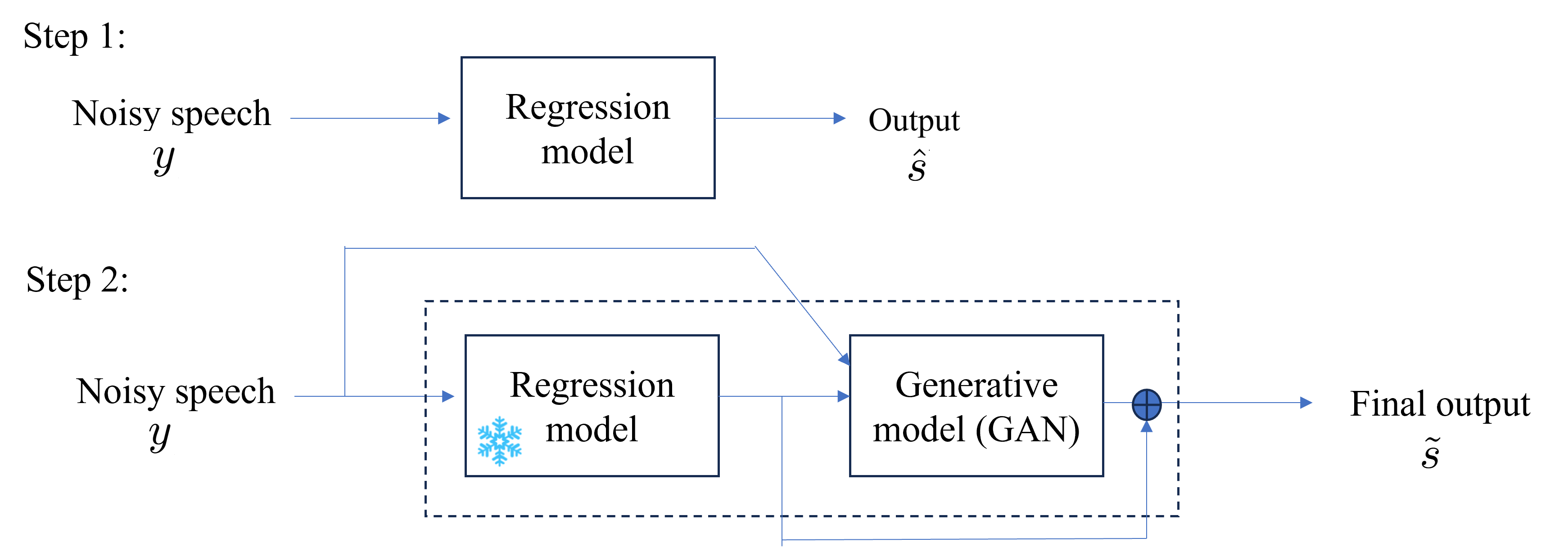}
\end{center}
\caption{Motivated by the distortion–perception tradeoff theory, the proposed two-stage framework integrates a frozen regression model with a residual generative model.}
\label{fig:R_and_G}
\end{figure}

According to the distortion–perception tradeoff theory~\citep{blau2018perception} shown in computer vision applications, speech restoration also faces a fundamental trade-off between \textbf{fidelity} (preserving linguistic content, speaker identity, emotion, and accent) and \textbf{perceptual quality}.
Generative and regression (also called discriminative) models tackle the problem in distinct ways. The regression model outputs the conditional expectation, $\hat{s} = E[S \mid Y = y]$, whereas the generative model produces samples from the conditional distribution, $\hat{s} \sim p(s \mid y)$, where $y$ corresponds to the degraded input speech signal. 
Under severe degradation like packet loss, bandwidth limitation, and low SNR, $y$ contains little information about the true clean signal so that
the regression output is biased towards prior mean $E[S \mid Y = y] \approx E[S]$. If the prior is multimodal (e.g., many plausible phonemes), the mean may be a blend of modes, which sounds muffled and unnatural, resulting in \textbf{over-smoothing problem}~\citep{ren2022revisiting, chao2025universal}. Conversely, in such a scenario, the generative model produces outputs drawn from the prior distribution, $p(s \mid y) \approx p(s)$, enabling the generation of natural-sounding speech consistent with the prior. Nevertheless, the lexical/linguistic content or speaker characteristics are not guaranteed to match the original signal, resulting in \textbf{hallucination problem} as defined in~\citep{ saijo2025interspeech, scheibler2024universal}. In summary, when $y$ is less informative, the regression model excels at preserving fidelity but may fail to improve quality, whereas the generative model enhances quality but often struggles to preserve fidelity. 
Specifically, posterior sampling from the posterior $p(s \mid y)$ results in a mean squared error (MSE) that is twice the minimum MSE (MMSE), which is achieved by the regression model~\citep{blau2018perception}.

We address this trade-off by leveraging insights from \cite{freirich2021theory}, which suggest that an optimal distortion-perception balance---minimizing MSE while satisfying the constraint of perfect perception---can be achieved by \textbf{optimally transporting} the posterior mean (MMSE estimate) toward the true data distribution (the derivation is provided in Appendix~\ref{Ap1}).
We further argue that \textbf{preserving fidelity, particularly linguistic content, is usually paramount} in applications of universal speech enhancement.
Therefore, we propose a sequential strategy that first uses a regression model to estimate the posterior mean, and then applies a generative model to correct only the over-smoothed regions through optimal transport (see Figure~\ref{fig:R_and_G}): 
\begin{enumerate}
    \item \textbf{Regression Stage:} We first train a regression model to convergence to estimate the posterior mean ($\hat{s} = E[S \mid Y = y]$), maximizing fidelity. We then freeze its weights.
    \item \textbf{Generative Stage:} We then employ a generative model to optimally transport the posterior mean toward the true data distribution by minimizing the Wasserstein distance ($W_2(p_{\hat{s}}, p_s)$).
    Crucially, we employ a residual connection between the regression model output and the final output, forcing the generative model to focus primarily on regions where characteristics diverge from those of real data to restore perceptual details.
\end{enumerate}

Previous work approximates the optimal transport using flow matching and achieves improved MSE in image restoration~\citep{ohayon2024posterior}. In addition to speech enhancement ~\citep{serbest2025deepfiltergan}, SEStream~\citep{huang2023two} and StoRM~\citep{lemercier2023storm} also achieve strong results in codec compression and dereverberation, respectively, by using a regression model to provide an initial prediction for a subsequent generative model. In this study, we leverage GAN-based methods to approximate optimal transport. Wasserstein GANs~\citep{arjovsky2017wasserstein} optimize an objective equivalent to the Wasserstein-1 distance between the source and target distributions, a principled metric from optimal transport theory.  Furthermore, as future work toward adapting the model to real-time scenarios, the single forward-pass generation of GAN offers lower computational latency. 

Beyond the formal derivation in Appendix~\ref{Ap1}, we offer an alternative explanation of why GANs can \textbf{focus on correcting over-smoothed regions while leaving other parts unchanged}. When a convolutional neural network (CNN) is employed as the discriminator, each element of the final feature map (the layer before the averaging operation used to produce the final prediction) has only a limited receptive field. Assuming the discriminator is Lipschitz continuous (e.g., via spectral normalization~\citep{miyato2018spectral}), the following constraint holds:
\begin{equation}
\label{eq3}
\begin{aligned}
\| D^{(l)}(\tilde{s}) - D^{(l)}(s) \| \leq L \| \tilde{s} - s \|,  \quad \forall l
\end{aligned}
\end{equation}

where $D^{(l)}(.)$ is the $l$-th discriminator layer, $L$ is the Lipschitz constant, and $\tilde{s}$ is the final model output. Here, let us focus on the receptive field of one element in the final feature map. The distance between the receptive field $\tilde{s}$ and the corresponding clean speech $s$ serves as an \textbf{upper bound} on the left-hand side of Equation~\eqref{eq3}, which corresponds to the \textbf{feature-matching loss} used during generator training \cite{salimans2016improved}. 
When the model accurately predicts the clean target within the receptive field (i.e., $\lVert \tilde{s} - s \rVert \approx 0$), Equation~\eqref{eq3} ensures that the feature-matching loss does not contribute gradients for these regions. Therefore, the generative model can mainly focus on correcting the over-smoothed regions of the regression model output (see Figures \ref{fig: GAN correction1} and \ref{fig: GAN correction2} in the Appendix for spectrogram visualization). Consequently, this two-stage framework can keep the fidelity while improving speech quality.

\subsection{Trade-off Between Training Data Scale and Quality}
\label{Trade-off}

\begin{figure}[ht]
\begin{center}
\includegraphics[width=0.52\linewidth]{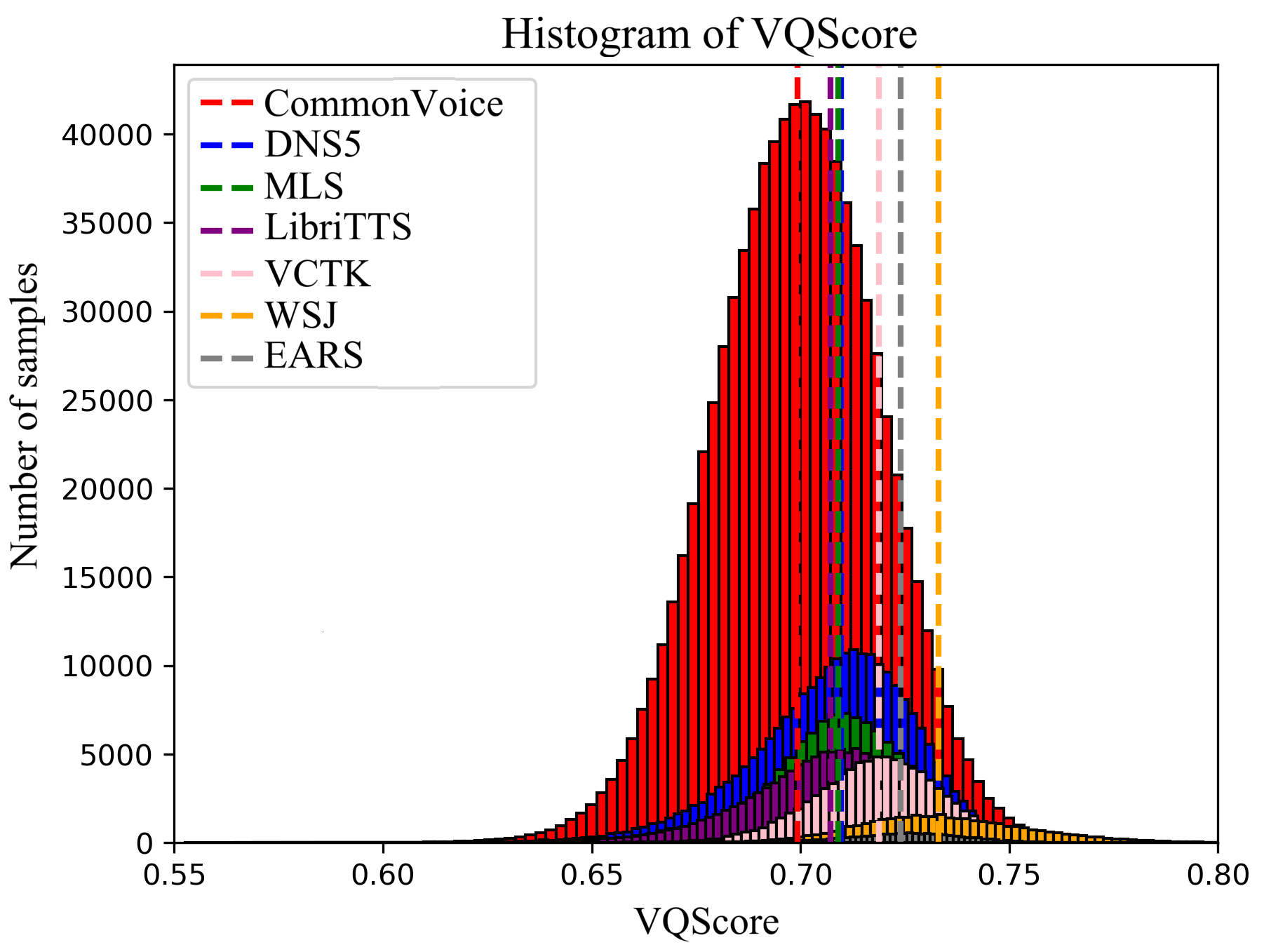}
\end{center}
\caption{Histogram of VQScore for URGENT 2025 Challenge Track 1 subsets. Dashed lines indicate median scores.}
\label{fig:Histogram}
\end{figure}

The URGENT 2025 Challenge (Track 1) provides approximately 2,500 hours of speech from diverse sources, including CommonVoice~\citep{ardila2019common}, DNS5~\citep{dubey2024icassp}, MLS~\citep{pratap2020mls}, LibriTTS~\citep{zen2019libritts}, VCTK~\citep{veaux2013voice}, WSJ~\citep{garofolo1993csr}, and EARS~\citep{richter2024ears} as summarized in Table~\ref{tab:dataset} (Appendix). Although the organizers have already filtered out non-speech samples using voice activity detection (VAD) and removed noisy samples based on the DNSMOS score, many recordings with audible background noise remain there~\citep{ saijo2025interspeech}. Leveraging its high correlation with subjective scores~\citep{zhang2025lessons} and fast inference capability (processing 2,500 hours of speech in less than 8 hours on a single NVIDIA A100 GPU), we employ VQScore~\citep{fu2024self} to analyze the quality distribution of each training data source. Figure~\ref{fig:Histogram} illustrates the VQScore distribution for each speech source (individual source histograms are provided in Figure \ref{fig: data_quality} in the Appendix). 

We observe a clear quality hierarchy: CommonVoice, the largest subset (1,300 hours), exhibits the lowest quality due to its crowdsourced nature.
Conversely, datasets like WSJ, EARS, and VCTK show consistently top-3 highest quality. 
Manual inspection reveals that low-VQScore samples often contain stationary background noise or entirely non-speech artifacts (examples are provided in the supplementary material).
These training targets can confuse the speech enhancement model and degrade its performance. 
To mitigate this, we apply a VQScore threshold to \textbf{further} filter low-quality samples. Finally, we leverage the high quality of the EARS dataset for a final fine-tuning stage. (Section~\ref{unseen}). Note that some extremely expressive EARS samples (e.g., whispering) receive low VQScore ratings, however, the dataset remains the cleanest overall source available.

\section{Experiments}
\label{headings}

\subsection{Dataset}
As noted earlier, the URGENT 2025 Challenge training dataset comprises multi-condition speech recordings across five languages (English, German, French, Spanish, and Chinese) with diverse sampling frequencies (8, 16, 22.05, 24, 32, 44.1, and 48 kHz), along with noise samples and RIRs. Seven types of distortions are considered: additive noise, reverberation, clipping, bandwidth limitation, codec artifacts, packet loss, and wind noise. We followed the organizers’ guidelines to simulate the validation set using the validation splits of the corpora~\citep{saijo2025interspeech}. The non-blind test set of URGENT 2025, consisting of 1,000 utterances, is used for evaluation.

\subsection{Model Architecture}
To enable a single model to operate across different sampling rates, we employ sampling frequency-independent (SFI) STFT~\citep{zhang2023toward}, which dynamically adjusts the FFT window and hop size according to the input sampling rate, ensuring a fixed time duration and consistent feature frame length across all sampling rates. To ensure that inputs with different sampling rates yield an integer number of frequency bins, we set the FFT window size to 320 points for the 8 kHz case 
(corresponding to a 40 ms window for all sampling rates).

Since the model architecture is not the primary focus of this paper, we adopt USEMamba with 30 layers~\citep{chao2025universal, chao2024investigation} as the regression model. USEMamba alternates between two types of sequence modeling modules (i.e., Mamba) applied to frequency features and time features in the time-frequency domain. For GAN training, we use a 6-layer USEMamba as the generator and CNN-based discriminators. 

To account for distinct feature patterns across frequency bands and support speech with varying sampling rates, we propose an \textbf{adaptive multi-band discriminator}, inspired by \cite{kumar2023high}. For each band corresponding to the input sampling rate (e.g., for 8 kHz, a single sub-band from 0-4 kHz; for 22.05 kHz, three sub-bands: 0-4 kHz, 4-8 kHz, and 8-11.025 kHz), we use a 5-layer 2-D convolutional network for local feature extraction. Sub-band features are concatenated along the frequency axis and passed through a final 2-layer 2-D convolution followed by global average pooling to produce the discriminator output. All models are trained on 8 NVIDIA A100 GPUs with a batch size of 1, allowing longer utterances to be processed without memory issues. We use AdamW with a learning rate of 0.0002 for the regression model, generator, and discriminator. The code will be released upon acceptance to facilitate reproducibility.

\subsection{Evaluation Metrics}
\label{metrics}

To address the dual objectives of improving perceptual quality and maintaining signal fidelity, we employ a comprehensive suite of metrics.
For standard reference-based evaluation, we report Perceptual Evaluation of Speech Quality (PESQ) for perceptual quality~\citep{Rix2001}, Extended Short-Time Objective Intelligibility (ESTOI) for intelligibility~\citep{jensen2016stoi}, and Signal-to-Distortion Ratio (SDR) for time-domain waveform distortion~\citep{roux2019sisdr}. Spectral deviation is further assessed using Mel Cepstral Distortion (MCD) and Log-Spectral Distance (LSD).

We also evaluate performance on downstream tasks using both task-independent and task-dependent measures. Task-independent metrics include SpeechBERTScore (SBERT)~\citep{saeki2024speechbertscore}, which utilizes self-supervised models to quantify enhancement quality, and Levenshtein Phoneme Similarity (LPS)~\citep{pirklbauer2023evaluation} for phoneme sequence similarity. Task-dependent metrics include speaker similarity (SpkSim) to measure speaker identity preservation and character accuracy (CAcc) to reflect ASR performance ~\cite{peng2024owsm}. Finally, we report non-intrusive metrics, including DNSMOS~\citep{reddy2022dnsmos}, NISQA~\citep{mittag2021nisqa}, and UTMOS~\citep{saeki2022utmos}, which estimate perceptual quality without requiring a clean reference.


Most Metrics (excluding non-intrusive ones and CAcc) require a clean speech reference. However, for the speech dereverberation task, the organizers provide \textbf{early-reflected speech} as the clean reference~\citep{saijo2025interspeech}. This creates a definition mismatch, as our model is trained on time-shifted anechoic clean speech. \textbf{Consequently, this discrepancy may penalize our leaderboard scores on reference-based metrics such as PESQ, ESTOI, SDR, MCD, LSD, SBERT, LPS, and SpkSim}, underestimating performance compared to the actual perceptual quality and fidelity obtained when using the original anechoic speech as the reference, as reported in Table~\ref{tab:shifted_anechoic} in the Appendix.

\subsection{Results on Training Data Filtering Based on Quality Estimation}

\begin{figure*}[t]
\centering
\subfloat[Different VQScore thresholds]{
    \includegraphics[width=0.48\textwidth]{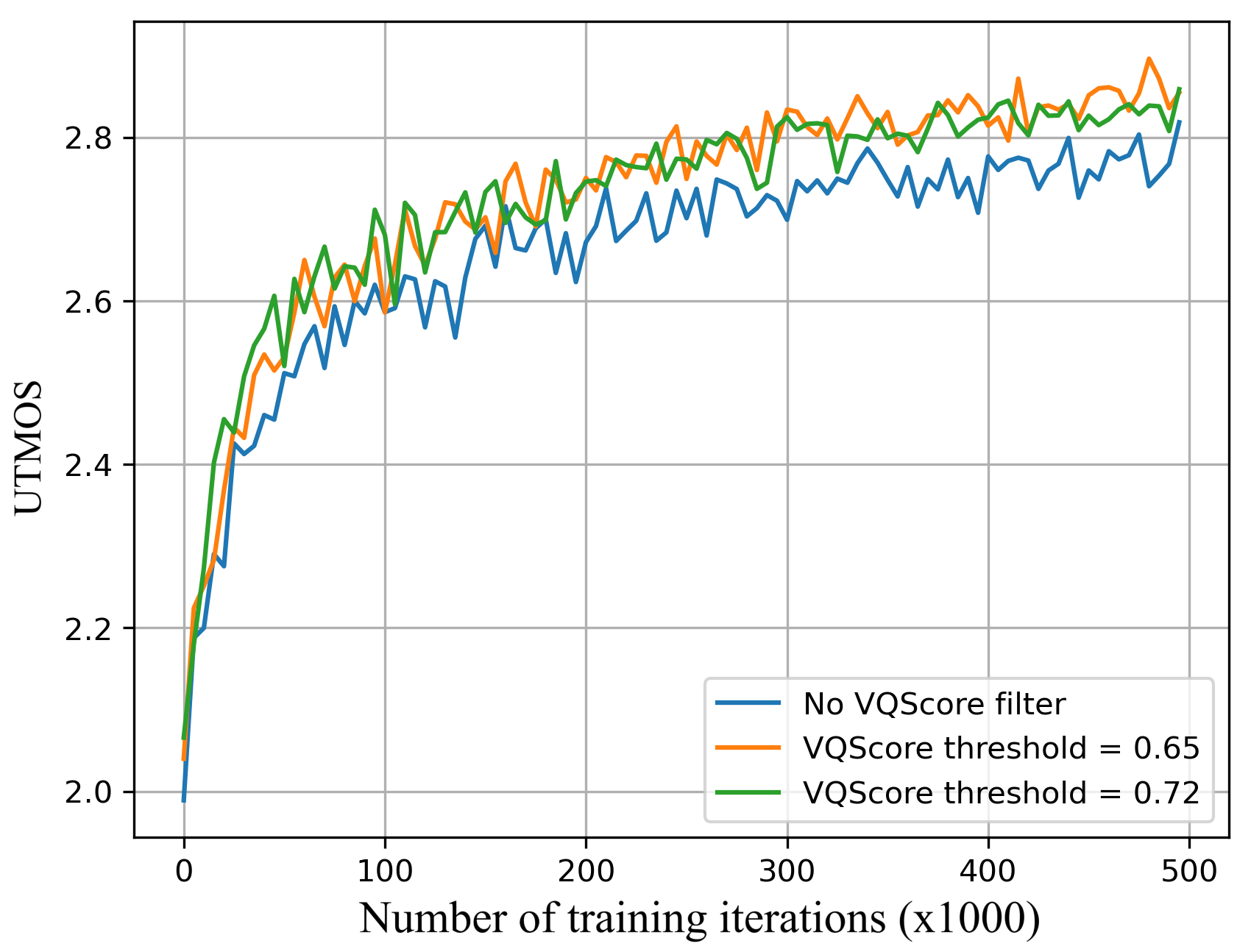}
}
\hfill
\subfloat[Different learning targets]{
    \includegraphics[width=0.48\textwidth]{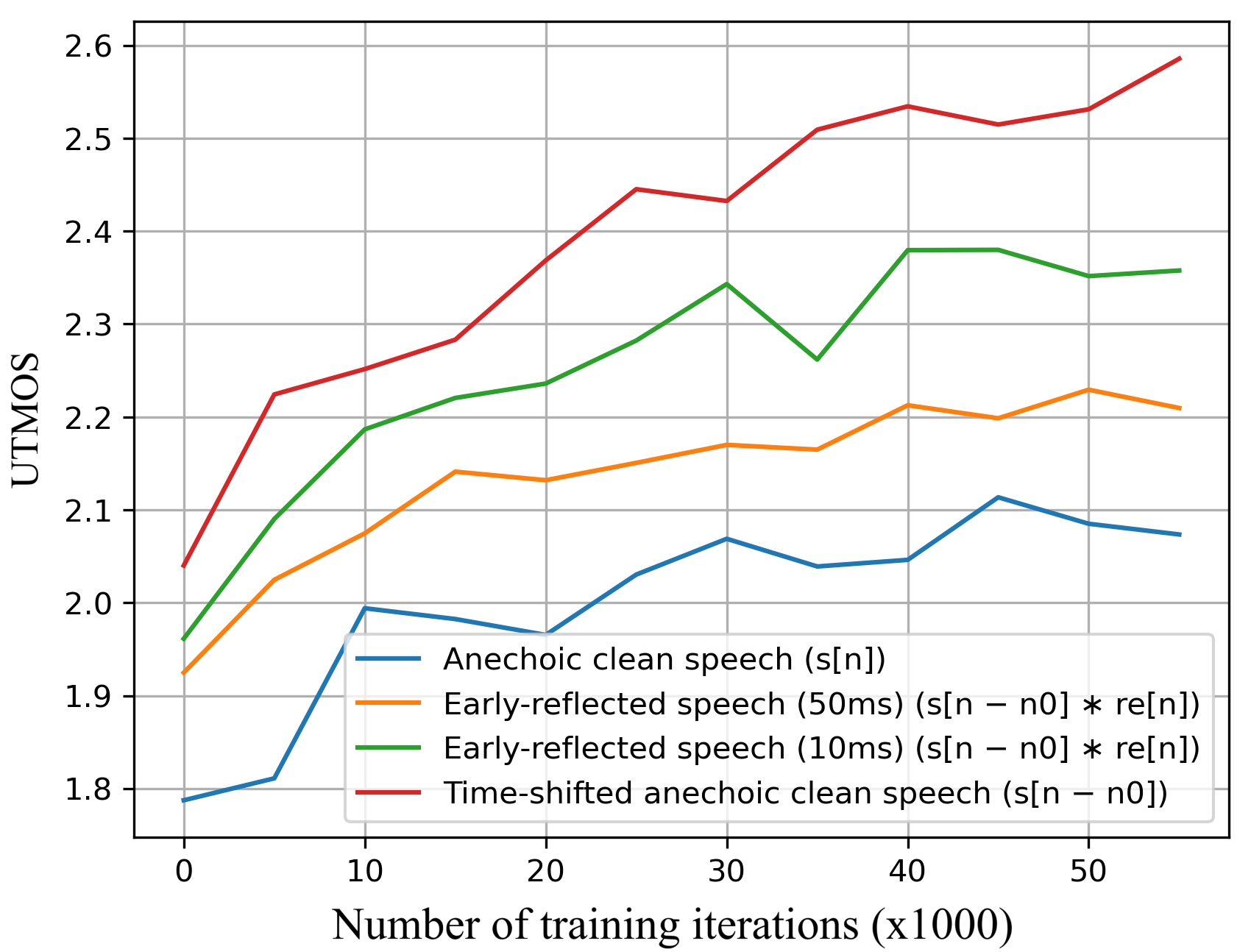}
}

\caption{Learning curves of UTMOS scores on the validation set under (a) different VQScore filtering thresholds and (b) different learning targets.}
\label{fig:UTMos_VqScore}
\end{figure*}

Based on the observations in Section~\ref{Trade-off}, we first determine an appropriate VQScore threshold, such that training samples with scores below the threshold are excluded. We consider three thresholds: 0.50 (no filtering), 0.65, and 0.72, corresponding to 2,518 (original size), 2,506, and 629 hours of training data, respectively. We then train three models on these datasets using time-shifted anechoic clean speech as targets and plot the UTMOS learning curves on the validation set in Figure~\ref{fig:UTMos_VqScore}\mbox{(a)}. Without VQScore filtering (\textcolor{Cerulean}{blue line}), the model performs the worst, \textbf{even though the Challenge organizers already removed speech samples with DNSMOS scores below 3}~\citep{saijo2025interspeech}. With a threshold of 0.72 (\textcolor{LimeGreen}{green line}), the model initially achieves the best performance in the early stage of training due to the higher data quality, but later lags behind the model trained with a threshold of 0.65 (\textcolor{BurntOrange}{orange line}), likely due to the reduced data volumn. Based on these results, 
we adopt a threshold of 0.65 in subsequent experiments, as it provides a good balance between training data quality and quantity.

\subsection{Results of Applying Time-Shifted Anechoic Clean Speech as Learning Targets}

We first examine the UTMOS learning curves under different training targets, shown in Figure~\ref{fig:UTMos_VqScore}\mbox{(b)}. As discussed in Section~\ref{anechoic clean}, directly using the original anechoic clean speech $s[n]$ (\textcolor{Cerulean}{blue line}) as the learning target yields the worst performance, consistent with previous studies~\citep{valin2022dereverb, zhao2020monaural}. Therefore, we exclude this learning target from subsequent evaluations. 

Table~\ref{tab:track1_results} reports the non-blind test results of our proposed approach, which uses time-shifted anechoic clean speech $s[n-n_0]$ as the dereverberation target, alongside the baseline TF-GridNet~\citep{wang2023tf} and the top three ranked systems in the challenge 
(Rank 1: Team Bobbsun~\citep{sun2025scaling}, Rank 2: Team rc~\citep{chao2025universal}, Rank 3: Team Xiaobin~\citep{rong2025ts}). 
The full leaderboard is publicly available \footnote{https://urgent-challenge.github.io/urgent2025/leaderboard/} and as presented in Table~\ref{tab:challenge_results} in the Appendix. Compared with the early-reflected target, replacing it with time-shifted anechoic clean targets yields substantial improvements on non-intrusive quality metrics (DNSMOS from 3.06 to 3.25, NISQA from 3.23 to 3.85, and UTMOS from 2.26 to 2.76) and ASR performance (CAcc from 87.62 to 89.41). The improvement in CAcc indicates that these quality gains do not come at the expense of hallucination. 

As noted in Section~\ref{metrics}, the intrusive and downstream metrics on the official leaderboard may not accurately reflect the performance of models trained with shifted anechoic speech targets, because the evaluation uses early-reflected speech as the reference. To correct for this, Table~\ref{tab:shifted_anechoic} (Appendix) presents the same metrics computed using \textbf{the original anechoic clean speech as the reference}. Under this consistent definition, shifted anechoic speech targets significantly outperform early-reflected ones. These findings indicate that using early-reflected speech as the learning target still degrades the output quality of USE models, leading to enhanced audio that remains noticeably \textbf{reverberant}. Spectrogram comparison in Figure~\ref{fig: shifted anechoic clean speech VS early-reflected} (Appendix) and audio examples on our demo page further illustrate this effect.

\begin{table*}[t]
    \centering
    \caption{Non-blind test set results of the URGENT 2025 Challenge. All metrics are ``higher is better'', except MCD and LSD. Rank $N$ denotes the system ranked $N^{th}$ in the challenge. Note that shaded metrics are not directly comparable across the two learning targets due to mismatches in the definition of the “clean” reference. See Tables~\ref{tab:data_filter} to ~\ref{tab:two_stage_gan} in the Appendix for the ablation studies using the original anechoic clean speech as the reference.}
    \resizebox{0.99\textwidth}{!}{%
    \begin{tabular}{lcccccccccccccc}
        \toprule
        \textbf{Team /}
        & \multicolumn{3}{c}{\textbf{Non-intrusive SE metrics}}
        & \multicolumn{5}{c}{\textbf{Intrusive SE metrics}} 
        & \multicolumn{2}{c}{\textbf{Task-ind.}} 
        & \multicolumn{2}{c}{\textbf{Task-dep.}} 
        \\
        \cmidrule(lr){2-4}
        \cmidrule(lr){5-9}
        \cmidrule(lr){10-11}
        \cmidrule(lr){12-13}
        
        \textbf{Rank} & 
        \textbf{DNSMOS} & \textbf{NISQA} & \textbf{UTMOS} &
        \ccc{\textbf{PESQ}} & \ccc{\textbf{ESTOI}} & \ccc{\textbf{SDR}} & \ccc{\textbf{MCD} $\downarrow$} & \ccc{\textbf{LSD $\downarrow$}} &  \ccc{\textbf{SBERT}} & \ccc{\textbf{LPS}} & \ccc{\textbf{SpkSim}} & \textbf{CAcc} \\
        \midrule
        
        Noisy & 1.84 & 1.69 &	1.56 &	1.37 &	0.61 &	2.53 &	7.92 &	5.51 &	0.75 &	0.62 &	0.63 &	81.29 \\
        
        Baseline 
        & 2.94 &	2.89 &	2.11 &	2.43 &	0.80 &	11.29 &	3.32 &	2.85 &	0.86 &	0.79 &	0.80 &	84.96 \\

        Rank 3
        & 3.00 &	3.45 &	2.31 &	2.74 &	0.84 &	13.06 &	3.30 &	3.08 &	0.89 &	0.84 &	0.83 &	87.94 \\

        Rank 2
        & 3.01 & 3.21 &	2.30 &	2.79 &	0.85 &	13.11 &	2.93 &	2.94 &	0.90 &	0.85 &	0.84 &	88.05 \\

        Rank 1
        & 3.01 & 3.41 	& 2.40 & \textbf{2.95} & \textbf{0.86} 	& \textbf{14.33} & 3.01 & 2.83 &	\textbf{0.91} & \textbf{0.86} &	0.85 &	88.92 \\ 
        \bottomrule
        



        \textbf{Early reflected} & 3.06 & 3.23 & 2.26 & 2.81 & 0.85 & 12.28 & \textbf{2.87} & \textbf{2.66} & 0.90 & 0.84  & 0.82  & 87.62  \\ 

        

        \textbf{ +GAN correction} & 3.04 & 3.53 & 2.30 & 2.78 & 0.84 & 12.25 & 2.97 & 2.75 & 0.90 & 0.84  & \textbf{0.85}  & 88.13  \\ 

        


        

        \textbf{Shifted anechoic} 
        & 3.25 & 3.85 & 2.76 & 2.41 & 0.77 & 8.23 & 3.63 & 3.28 & 0.89 & 0.84 & 0.82 & 89.41 \\

        \textbf{ +GAN correction} 
        & \textbf{3.26} & \textbf{4.12} & \textbf{2.80} & 2.38 & 0.76 & 8.18 & 3.73 & 3.51 & 0.89 & 0.84 & 0.83 & \textbf{89.88} \\
        

        
        
        
        \bottomrule
    \end{tabular}
    }
    \label{tab:track1_results}
\end{table*}

\subsection{Results of the Two-Stage Combination of Regression and Generative Models}

To verify that the two-stage framework enables the GAN to focus primarily on correcting over-smoothed regions while leaving well-predicted region intact, we compute the average 
correlation coefficient between two magnitude-residual spectrograms on the non-blind test set: the `clean–regression residual' (clean speech $s$ minus regression output $\hat{s}$) and the `final–regression residual' (final output $\tilde{s}$ minus regression output $\hat{s}$). We obtain a high correlation of 0.78, indicating that the GAN corrections are strongly aligned with the residual errors of the regression model and thus largely preserve signal fidelity (see Figures \ref{fig: GAN correction1} and \ref{fig: GAN correction2} in Appendix for spectrogram visualization).

From Table~\ref{tab:track1_results}, we observe that applying GAN refinement to the regression output consistently and significantly improves NISQA, SpkSim, and CAcc, with moderate gains in UTMOS, while having only marginal or negligible impact on most intrusive SE metrics. This suggests that fidelity is well preserved while perceptual quality and downstream performance are enhanced. The overall ranking of the non-blind test set is summarized in Table~\ref{tab:challenge_results} in the Appendix, where \textbf{GAN correction leads to an improved leaderboard rank}. To further compare our two-stage framework with other GAN training paradigm, Figure~\ref{fig: direct fine-tuning vs two-stage GAN} (Appendix) plots the validation learning curves for a convention approach (pre-training with a regression loss followed by adversarial fine-tuning) versus our two-stage GAN correction. Across training, our two-stage framework consistently achieves lower magnitude, phase, and time-domain losses, as well as higher PESQ scores.
\textbf{In summary, the shifted anechoic target combined with GAN correction attains state-of-the-art performance across all non-intrusive metrics and ASR CAcc on the URGENT 2025 non-blind test set.} Ablation studies evaluating the effects of the three proposed techniques, with the original anechoic clean speech as the reference, are reported in Appendix~\ref{Ap2}. 

\subsection{Subjective Evaluation}

To subjectively evaluate audio quality under different USE settings, we conducted a listening test using a 5-point scale. A total of 23 participants listened to the audio samples and rated them from 1 (poor quality) to 5 (excellent quality).
Ten English utterances were randomly selected for each condition: noisy input, USE using early reflected target, shifted anechoic target, and shifted anechoic target + GAN. The average quality scores for these conditions are 1.86, 3.00, 3.68, and 3.97, respectively, with 95\% confidence intervals of 0.19, 0.26, 0.20, and 0.19. These results verify the effectiveness of the proposed method and are consistent with the trends observed in the corresponding objective metrics. Some audio samples are available at \url{https://jasonswfu.github.io/USE/}.














\subsection{Comparison with Other Open-Source USE Models}

\begin{table*}[th]
    \centering
    \caption{Comparison with other open-source USE models on the subsets of the URGENT 2025 non-blind test set. All metrics are “higher is better,” except MCD and LSD. }
    \resizebox{0.99\textwidth}{!}{%
    \begin{tabular}{lcccccccccccccc}
        \toprule
        \textbf{Team /}
        & \multicolumn{3}{c}{\textbf{Non-intrusive SE metrics}}
        & \multicolumn{5}{c}{\textbf{Intrusive SE metrics}} 
        & \multicolumn{2}{c}{\textbf{Task-ind.}} 
        & \multicolumn{2}{c}{\textbf{Task-dep.}} 
        \\
        \cmidrule(lr){2-4}
        \cmidrule(lr){5-9}
        \cmidrule(lr){10-11}
        \cmidrule(lr){12-13}
        
        \textbf{Rank} & 
        \textbf{DNSMOS} & \textbf{NISQA} & \textbf{UTMOS} &
        \ccc{\textbf{PESQ}} & \ccc{\textbf{ESTOI}} & \ccc{\textbf{SDR}} & \ccc{\textbf{MCD} $\downarrow$} & \ccc{\textbf{LSD $\downarrow$}} &  \ccc{\textbf{SBERT}} & \ccc{\textbf{LPS}} & \ccc{\textbf{SpkSim}} & \textbf{CAcc} \\
        \midrule
        
         &  &  &	 &	 &	 &	48k Hz &	 &	 &	 &	 &	 &	 \\
        \midrule

        Noisy & 2.04 & 1.83 & 1.99 & 1.28 &	0.56 & 2.29 & 7.77 & 5.50 & 0.78 &	0.77 & 0.69 & 90.60 \\

        ClearerVoice
        & 2.97 &	3.38 &	3.02 &	2.09 &	0.72 &	11.55 &	5.08 &	5.15 &	0.85 &	0.87 &	0.63 &	89.90 \\

        \textbf{Proposed} 
        & \textbf{3.31} & \textbf{4.41} &	\textbf{3.55} & \textbf{2.65} &	\textbf{0.77} &	\textbf{12.79} & \textbf{3.93} & \textbf{3.19} &	\textbf{0.89} &	\textbf{0.91} &	\textbf{0.87} &	\textbf{92.50} \\

        \bottomrule
        
        &  &  &  &  & & 44.1k Hz&  & &  &   &   &     \\ %
        
        \midrule

        Noisy & 1.91 & 1.79 & 1.52 & 1.33 &	0.64 & 3.34 & 7.44 & 5.62 & 0.76 &	0.63 &	0.71 &	83.60 \\
        
        Resemble Enhance & 3.13 & 3.68 & 2.11 & 1.33 & 0.45 & -15.01 & 11.02 & 7.93 & 0.69 & 0.47  & 0.61  & 47.20 \\ %

        \textbf{Proposed} 
        & \textbf{3.32} & \textbf{4.15} & \textbf{2.68} & \textbf{2.28} & \textbf{0.78} & \textbf{7.18} & \textbf{3.91} & \textbf{3.47} & \textbf{0.90} & \textbf{0.85} & \textbf{0.88} & \textbf{92.20} \\

        \bottomrule
    \end{tabular}
    }
\label{tab:open_source_results}
\end{table*}

We next compare our proposed USE model (shifted anechoic target + GAN correction) with other popular open-source USE models, ClearerVoice-Studio~\citep{zhao2025clearervoice}\footnote{https://github.com/modelscope/ClearerVoice-Studio}
 and Resemble Enhance\footnote{https://github.com/resemble-ai/resemble-enhance}
. Although this comparison is not strictly controlled due to differing training data, it still offers useful insight into the relative strengths and limitations of existing approaches. Since ClearerVoice-Studio supports only 16 kHz and 48 kHz inputs, and Resemble Enhance only 44.1 kHz, we evaluate on the corresponding subsets of the URGENT 2025 non-blind test set, as reported in Table~\ref{tab:open_source_results}.
Our proposed method outperforms ClearerVoice-Studio, a regression-based model built on MossFormer2~\citep{zhao2024mossformer2}, although ClearerVoice-Studio still achieves reasonable enhancement quality. Resemble Enhance, based on latent conditional flow matching, substantially improves non-intrusive quality metrics but yields low intrusive scores and CAcc, suggesting a tendency to hallucinate content, which is consistent with prior findings on purely generative models~\citep{saijo2025interspeech}.  

\subsection{Evaluation on Unseen Languages}
\label{unseen}

\begin{table}[t]
  \caption{Speech enhancement results for unseen languages from the FLEURS dataset.}
  \label{table:unseen_language}
  \centering
  \small
  \renewcommand{\arraystretch}{0.7} 
  \setlength{\extrarowheight}{-1pt} 
  \begin{tabular}{cccccc}
    \hline
      &  \makecell{\textbf{DNSMOS}} & 
      \makecell{\textbf{SpkSim}} &
       \makecell{\textbf{CAcc}} 
      \\
      
    \midrule
    
      & Italian (it$\_$it)  &  &  &    \\
    \midrule
    Original & 3.12 
    & - & 97.28 \\
    FLEURS-R 
    & \textbf{3.37} 
    & 0.87 & 97.69\\
    Proposed  & 3.20 
    & 
    \textbf{0.98} & 97.00 \\
    Proposed (EARS)   & 3.27 
    & 
    0.97 & \textbf{98.09} \\
        \midrule
    & Dutch (nl$\_$nl) &  &  &    \\
    \midrule
    Original & 2.99 
    & - & \textbf{97.40} \\
    FLEURS-R 
    & \textbf{3.36} 
    & 0.88 & 97.19 \\
    Proposed  & 3.13 
    & 
    \textbf{0.97} & 97.18 \\
    Proposed (EARS)   & 3.28 
    & 
    0.95 & 97.26 \\
        \midrule
    & Japanese (ja$\_$jp) &  &  &    \\
    \midrule
    Original & 2.96 
    & - & 95.34 \\
    FLEURS-R 
    & \textbf{3.36} 
    & 0.88 & 95.14 \\
    Proposed  & 3.07 
    & 
    \textbf{0.98} & 95.30 \\
    Proposed (EARS)   & 3.18 
    & 
    0.95 & \textbf{95.43} \\
    \bottomrule
  \end{tabular}
\end{table}

One emerging application of USE is training data cleaning for downstream speech generative models (e.g., text-to-speech)~\citep{koizumi2023miipher, karita2025miipher, koizumi2023libritts, ma2024fleurs}. This is particularly important for low-resource languages, where studio-quality recordings are scarce. To support this use case, the USE model must be \textbf{language-agnostic}. \cite{saijo2025interspeech} report that regression models are relatively insensitive to language variations, whereas purely generative models (e.g., latent diffusion and neural vocoders) tend to be more language dependent. This observation also motivates our two-stage design, where the generative model is used only to refine over-smoothed regions of a regression model’s output.

Following Miipher-2~\citep{karita2025miipher}, we use the FLEURS dataset~\citep{conneau2023fleurs} to evaluate model performance on unseen languages. We select three languages (Italian, Dutch, and Japanese) and assess speech quality with DNSMOS, speaker similarity (relative to the original)~\citep{desplanques2020ecapa}, and ASR CAcc~\citep{radford2023robust} before and after different speech restoration, as summarized in Table~\ref{table:unseen_language}. We also include FLEURS-R~\citep{ma2024fleurs}, a restored version of FLEURS processed by Miipher-2, obtained from the official release\footnote{https://huggingface.co/datasets/google/fleurs-r}. Miipher-2 uses acoustic features extracted by the Universal Speech Model~\citep{zhang2023google}, pre-trained on 12 million hours of speech across more than 300 languages, and reconstructs back with the WaveFit vocoder~\citep{koizumi2023wavefit}.

We observe that some FLEURS samples contain only mild stationary wideband noise or electrical microphone hiss (see Figure~\ref{FLEURS} in the Appendix), artifacts that still appear in our training data even after VQScore filtering. Our original model (shifted anechoic target + GAN) does not fully remove such low-level background noise, likely because some ``clean'' training examples remain imperfect and the model learns to preserve these subtle distortions. To address this, we fine-tune the model solely on the highest-quality subset of our training data (EARS), and denote the result as Proposed (EARS) in Table~\ref{table:unseen_language}. Although FLEURS-R achieves slightly higher DNSMOS, its speaker characteristics deviate more from the original speech compared to our proposed methods, highlighting the fidelity-quality trade-off. Namely, some generative restoration may enhance perceptual quality at the expense of speaker similarity. Between our two proposed variants, Proposed (EARS) provides better speech quality and ASR accuracy, reinforcing the importance of training data quality for USE.

\subsection{Application to TTS Training Data Cleaning}

\begin{table*}[t]
    \caption{Zero-shot TTS evaluation after training data cleaning using our USE model on unseen languages. We report the 95\% confidence intervals based on standard errors calculated from 10 independent runs per dataset.}
    \centering
    \small
    \setlength{\tabcolsep}{4pt} 
    \resizebox{\textwidth}{!}{
    \begin{tabular}{cccrccc}
        \toprule
        \textbf{Language} & \textbf{Context audio} & \textbf{Train audio} & \textbf{CER (\%)} & \textbf{WER (\%)} & \textbf{SpkSim} & \textbf{FCD} \\
        \midrule
        {Dutch} 
            & original  & original  & 14.28 $\pm$ 0.98 & 19.60 $\pm$ 0.76 & 0.6064 $\pm$ 0.0080 & 0.2444 $\pm$ 0.0155 \\
            & enhanced  & enhanced  & \textbf{7.75} $\pm$ 0.83 & \textbf{13.66} $\pm$ 0.71 & \textbf{0.6603} $\pm$ 0.0047 & \textbf{0.1837} $\pm$ 0.0086 \\
        \midrule
        {Italian} 
            & original  & original  & 11.13 $\pm$ 0.94 & 19.20 $\pm$ 0.94 & 0.6004 $\pm$ 0.0034 & 0.1846 $\pm$ 0.0042 \\
            & enhanced  & enhanced  & \textbf{8.30} $\pm$ 0.52 & \textbf{15.98} $\pm$ 0.53 & \textbf{0.6006} $\pm$ 0.0032 & \textbf{0.1373} $\pm$ 0.0021 \\
        \bottomrule
    \end{tabular}
    }
    \label{tab:tts}
\end{table*}

The scarcity of studio-quality data hinders progress in TTS modeling, particularly for low-resource languages. We investigate the hypothesis that TTS training data restored by our USE can alleviate these bottlenecks. We exclude high-resource English and experiment with a training dataset comprised of four European languages: French~(7.3k hours), German~(5.5k hours), Dutch~(760 hours), and Italian~(123 hours), sourced from CML-TTS~\citep{oliveira2023cml} and Emilia-YODAS~\citep{he2025emilia}, with a sampling rate of 22.05 kHz. 

We employ Zero-Shot Koel-TTS~\citep{hussain2025koel}, a state-of-the-art encoder-decoder Transformer TTS backbone. This model features an autoregressive decoder that generates speech tokens conditioned on a text transcript and a speaker audio prompt. The model contains approximately 378M parameters and operates on low-frame-rate~(21.5 FPS) audio codec tokens encoded by NanoCodec~\citep{casanova2025nanocodec}. Text transcripts are processed using a ByT5 byte-level tokenizer~\citep{xue2022byt5}.

We train the multilingual TTS model following the configurations of Koel-TTS~\citep{hussain2025koel}. We then evaluate the model on Dutch and Italian (unseen language during our USE training) using character error rate~(CER), word error rate~(WER), speaker similarity between context and generated audio, and Fr\'echet codec distance\footnote{Fr\'echet codec distance (FCD) adapts FID~\citep{heusel2017gans} to measure the distance between real and generated distributions in the codec's dequantized embedding space.}~(FCD). Table~\ref{tab:tts} demonstrates substantial improvements across all metrics when both context and training target audio are enhanced by our USE model. Furthermore, we observe that the TTS model consistently benefits when either the context or target audio is enhanced, as detailed in Table~\ref{tab:additional_tts} (Appendix). These experiments indicate that our USE model effectively unlocks the potential of existing large-scale, multilingual, noisy speech datasets for training high-quality TTS models.

\section{Conclusion}
This paper systematically investigates three critical challenges in developing USE models. 
First, we show that time-shifted anechoic clean speech is a better dereverberation target than conventional early-reflected speech, improving both perceptual quality and downstream ASR performance. Second, motivated by the distortion--perception trade-off theory, we propose a simple two-stage framework that balances fidelity and perceptual quality by combining a regression model with a residual generative refinement model, correcting over-smoothed regions without hallucinations. Third, we demonstrate that USE performance is strongly limited by training data quality: rigorous filtering and fine-tuning on the cleanest subset consistently yields better enhancement. Finally, our model generalizes well across languages while preserving high fidelity, making it effective for improving training data for downstream speech generation tasks.



\newpage
\bibliographystyle{unsrt}
\bibliography{neurips_conference}

\newpage
\appendix
\onecolumn
\section{Appendix}

\subsection{The Distortion-Perception Tradeoff}
\label{Ap1}

The performance of a speech enhancement model can be characterized by two criteria: 1) fidelity, measured by the average proximity of estimated speech $\tilde{s}$ to the clean speech $s$, and 2) perceptual quality, the degree to which the distribution of $\tilde{s}$
is close to that of $s$. To alleviate the hallucination problem in generative models, our goal is to \textbf{achieve minimal distortion under a given level of perceptual quality $P$}. Mathematically, we are dealing with
the distortion-perception (DP) function ~\citep{freirich2021theory},
\begin{equation}
D(P) = 
\min_{p_{\tilde{s}\mid y}}
\{
\;\mathbb{E}\!\left[d(s,\tilde{s})\right]
\quad
\text{s.t.}\quad
d_p\!\left(p_s, p_{\tilde{s}}\right) \le P 
\},
\label{eq4}
\end{equation}
Here, $d(.,.)$ denotes the distortion criterion and $d_p(.,.)$ denotes a divergence measure between two probability distributions. As shown in~\citep{blau2018perception}, $D(P)$ is \textbf{monotonically non-increasing and convex} under most commonly used divergence measures. Thus, traversing this function (or near this region) reveals a \textbf{trade-off between distortion and perceptual quality}.

Following ~\citep{freirich2021theory}, we consider squared-error distortion $d(s,\tilde{s}) = \lVert s - \tilde{s} \rVert_2^2$ and the Wasserstein
distance $d_p\!\left(p_s, p_{\tilde{s}}\right)
= W_2\!\left(p_s, p_{\tilde{s}}\right)$.
Then, Equation~\eqref{eq4} can be written as:
\begin{equation}
D(P) = 
\min_{p_{\tilde{s}\mid y}}
\{
\;\mathbb{E}\!\left[\lVert s - \tilde{s} \rVert_2^2\right]
\quad
\text{s.t.}\quad
W_2\!\left(p_s, p_{\tilde{s}}\right) \le P
\},
\label{eq5}
\end{equation}
Note that, without any constraints, the minimal distortion $D^*$ can be easily achieved by $s^* = \mathbb{E}[s \mid y]$.

If $\tilde{s}$ is independent of $s$ given $y$, the first term in Equation~\eqref{eq5} can be written as $\mathbb{E}\!\left[\lVert s - \tilde{s} \rVert^2\right]
=
\mathbb{E}\!\left[\lVert s - s^* \rVert^2\right]
+
\mathbb{E}\!\left[\lVert s^* - \tilde{s} \rVert^2\right] = D^* + \mathbb{E}\!\left[\lVert s^* - \tilde{s} \rVert^2\right]$. The DP function can therefore be rewritten as: 
\begin{equation}
D(P) = 
D^* + \min_{p_{\tilde{s}\mid y}}
\{
\;\mathbb{E}\!\left[\lVert s^* - \tilde{s} \rVert_2^2\right]
\quad
\text{s.t.}\quad
W_2\!\left(p_s, p_{\tilde{s}}\right) \le P \},
\label{eq6}
\end{equation}
The minimal distortion under perfect perceptual quality, denoted by $D(0)$ (i.e., $P$=0, which implies $W_2\!\left(p_s, p_{\tilde{s}}\right)=0$, and hence the constraint become $p_{\tilde{s}} = p_s$), is given by:
\begin{equation}
D(0) = 
D^* + \min_{p_{\tilde{s}\mid y}}
\{
\;\mathbb{E}\!\left[\lVert s^* - \tilde{s} \rVert_2^2\right]
\quad
\text{s.t.}\quad
p_{\tilde{s}} = p_s \},
\label{eq7}
\end{equation}

Since the objective ($\mathbb{E}\!\left[\lVert s^* - \tilde{s} \rVert_2^2\right]$) depends only on $p_{s^* \tilde{s}}$, we can rewrite the constraint as:

\begin{equation}
D(0) = 
D^* + \min_{p_{s^* \tilde{s}}}
\{
\;\mathbb{E}\!\left[\lVert s^* - \tilde{s} \rVert_2^2\right]
\quad
\text{s.t.}\quad
p_{s^* \tilde{s}}  \in \; \Pi(p_s, p_{s^*}) \},
\label{eq8}
\end{equation}

where $\Pi(p_s, p_{s^*})$ denotes the set of all joint distributions with marginals $p_s$ and $p_{s^\ast}$. Because the second term in Equation~\ref{eq8} corresponds to the Wasserstein distance between $p_s$ and $p_{s^*}$, we finally obtain:

\begin{equation}
D(0) = 
D^* +  W_2\!\left( p_{s^*}, p_s\right),
\label{eq9}
\end{equation}

Therefore, we can minimize the MSE while satisfying the perfect-perception constraint by \textbf{optimally transporting} the posterior mean prediction ($p_{s^\ast}$) to the real data distribution ($p_s$). Following this formulation, as illustrated in Figure~\ref{fig:R_and_G}, our two-stage training first uses a frozen regression model to estimate the posterior mean, and then employs the GAN generator to optimally transport this estimate toward the real data distribution by minimizing the Wasserstein distance. For a more detailed discussion of the properties of the DP function, please refer to ~\citep{blau2018perception} and ~\citep{freirich2021theory}.

\newpage

\subsection{Ablation study}
\label{Ap2}

In the following tables, we evaluate the effects of the three proposed techniques on USE performance using the \textbf{original anechoic clean speech} as the reference on the non-blind test set of the URGENT 2025 Challenge.

\begin{table*}[ht]
\centering
\caption{Comparison of speech enhancement performance with and without training data filtering.}
\label{tab:data_filter}
\resizebox{\textwidth}{!}{%
\begin{tabular}{lccccccccc}
\toprule
Method 
& DNSMOS & NISQA & UTMOS & PESQ & STOI & SBERT & LPS & SpkSim & CACC \\
\midrule
Unfiltered
& \textbf{3.06}
& 3.18
& 2.24
& 2.36
& 0.69
& 0.87
& 0.81
& 0.79
& 87.29 \\

Filtered
& \textbf{3.06}
& \textbf{3.24}
& \textbf{2.26}
& \textbf{2.40}
& \textbf{0.70}
& \textbf{0.88}
& \textbf{0.82}
& \textbf{0.80}
& \textbf{87.70} \\

\midrule
$p$-value
& 0.13
& 0.0007
& $2.36 \times 10^{-7}$
& $1.77 \times 10^{-11}$
& $4.23 \times 10^{-31}$
& $2.01 \times 10^{-8}$
& $5.24 \times 10^{-8}$
& $2.21 \times 10^{-6}$
& 0.46 \\
\bottomrule
\end{tabular}%
}
\end{table*}

\begin{table*}[ht]
\centering
\caption{Comparison between early-reflected and shifted anechoic training targets.}
\label{tab:shifted_anechoic}
\resizebox{\textwidth}{!}{%
\begin{tabular}{lccccccccc}
\toprule
Method 
& DNSMOS & NISQA & UTMOS & PESQ & STOI & SBERT & LPS & SpkSim & CACC \\
\midrule
Early-reflected
& 3.06
& 3.24
& 2.26
& 2.40
& 0.70
& 0.88
& 0.82
& 0.80
& 87.70 \\

Shifted anechoic
& \textbf{3.23}
& \textbf{3.82}
& \textbf{2.76}
& \textbf{2.70}
& \textbf{0.78}
& \textbf{0.89}
& \textbf{0.85}
& \textbf{0.81}
& \textbf{89.16} \\

\midrule
$p$-value
& $1.69 \times 10^{-76}$
& $6.89 \times 10^{-98}$
& $2.40 \times 10^{-107}$
& $2.34 \times 10^{-84}$
& $8.30 \times 10^{-103}$
& $1.59 \times 10^{-60}$
& $1.19 \times 10^{-46}$
& $2.82 \times 10^{-6}$
& $9.99 \times 10^{-6}$ \\
\bottomrule
\end{tabular}%
}
\end{table*}

\begin{table*}[ht]
\centering
\caption{Comparison between single-stage and two-stage training with GAN correction.}
\label{tab:two_stage_gan}
\resizebox{\textwidth}{!}{%
\begin{tabular}{lccccccccc}
\toprule
Method
& DNSMOS & NISQA & UTMOS & PESQ & STOI & SBERT & LPS & SpkSim & CACC \\
\midrule
Single-stage
& 3.23
& 3.82
& 2.76
& 2.70
& \textbf{0.78}
& \textbf{0.89}
& 0.85
& 0.81
& 89.16 \\

Two-stage (+ GAN)
& \textbf{3.26}
& \textbf{4.12}
& \textbf{2.80}
& \textbf{2.71}
& \textbf{0.78}
& \textbf{0.89}
& \textbf{0.86}
& \textbf{0.84}
& \textbf{89.88} \\

\midrule
$p$-value
& $3.83 \times 10^{-9}$
& $3.20 \times 10^{-68}$
& $9.58 \times 10^{-27}$
& $7.28 \times 10^{-60}$
& $4.59 \times 10^{-30}$
& $2.23 \times 10^{-10}$
& 0.25
& $2.40 \times 10^{-23}$
& 0.03 \\
\bottomrule
\end{tabular}%
}
\end{table*}

\subsection{Limitations}
We conducted several internal tests of this model, including on real noisy speech recorded with our laptops and cellphones, and it generally performs reasonably well. However, we do observe some failure cases. For example, in packet loss concealment, when the loss duration is too long, the reconstructed phonemes are often incorrect. Without additional conditioning (e.g., phoneme-based speech infilling or audio-visual speech enhancement using visual cues), and especially in multilingual settings, long-duration packet loss concealment remains a challenging task. Furthermore, due to its non-causal design, the model is not suitable for real-time applications; extending the framework to a causal setting is an important direction for future work.

\newpage

\begin{figure*}[h]
\begin{center}
\includegraphics[width=0.60\linewidth, height=0.4\linewidth ]{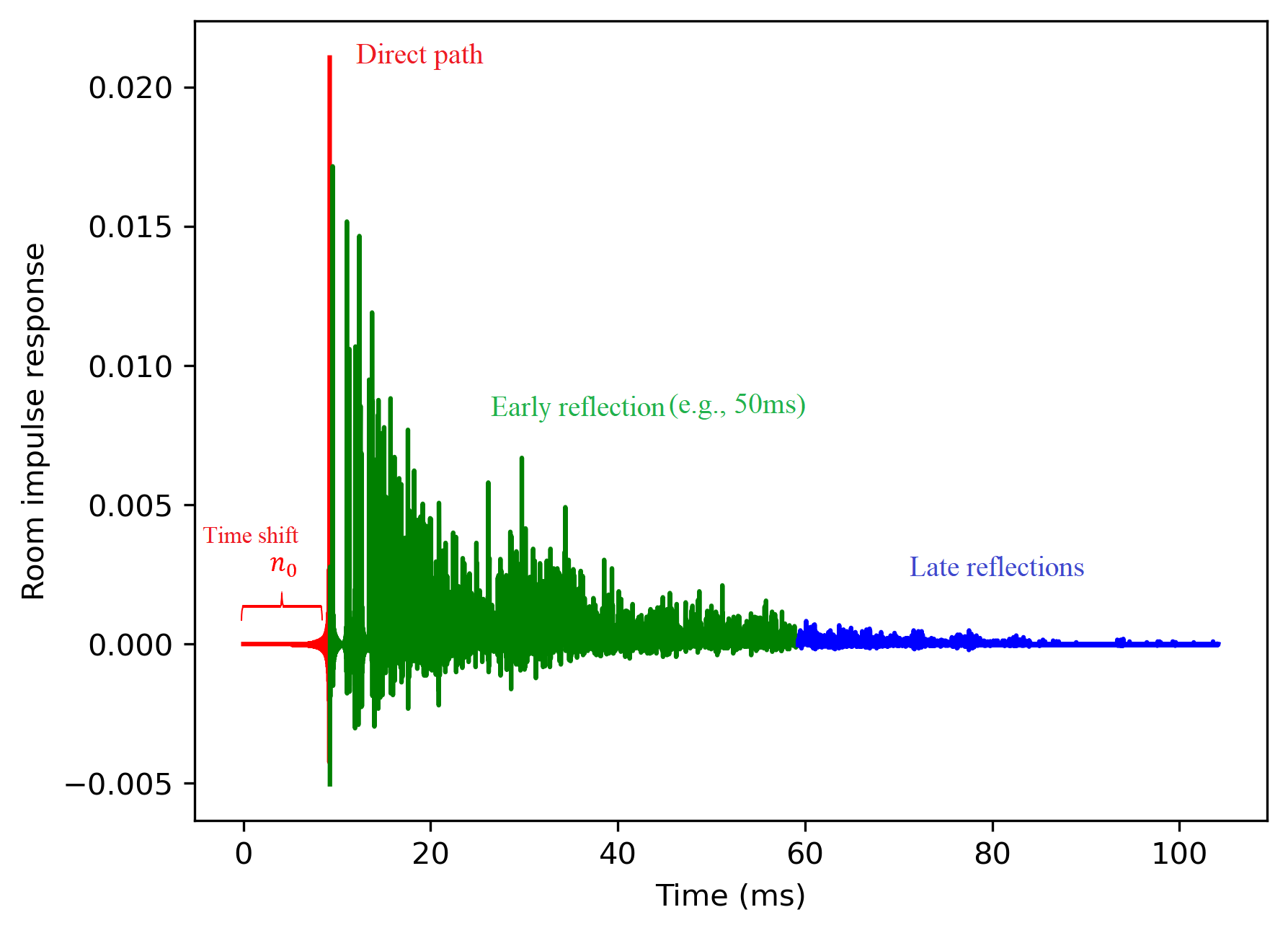}
\end{center}
\caption{An example of a room impulse response, highlighting the time shift 
$n_0$ introduced by the direct path.}
\label{fig:RIR}
\end{figure*}

\begin{table*}[h]
    \centering
    \caption{Dataset Composition for URGENT 2025 Challenge}
    \vspace{-0.6em}
    \resizebox{1\textwidth}{!}{%
    \begin{tabular}{lllcc}
        \toprule
        \textbf{Type} & \textbf{Corpus} & \textbf{Condition} & \textbf{Sampling (kHz)} & \textbf{Duration (h)} \\
        \midrule
        Speech
        & LibriVox (DNS5) & Audiobook & 8--48 & 350 \\
        & LibriTTS & Audiobook & 8--24 & 200 \\
        & VCTK & Newspaper-style & 48 & 80 \\
        & WSJ & WSJ news & 16 & 85 \\
        & EARS & Studio recording & 48 & 100 \\
        & Multilingual Librispeech (de, en, es, fr) & Audiobook & 8--48 & 450 \\
        & CommonVoice 19.0 (de, en, es, fr, zh-CN) & Crowd-sourced voices & 8--48 & 1300 \\
        \hline
        Noise 
        & AudioSet+FreeSound (DNS5) & Crowd-sourced + YouTube & 8--48 & 180 \\
        & WHAM! Noise & 4 urban environments & 48 & 70 \\
        & FSD50K (Filtered) & Crowd-sourced & 8--48 & 100 \\
        & Free Music Archive & Directed by WFMU & 8--44.1 & 200 \\
        \hline
        RIR
        & Simulated RIRs (DNS5) & SLR28 & 48 & 60k samples \\
        \bottomrule
    \end{tabular}
    }
    \label{tab:dataset}
\end{table*}

\begin{figure*}[t]
 \subfloat[Noisy]{\includegraphics[width=0.45\linewidth, height=4cm]{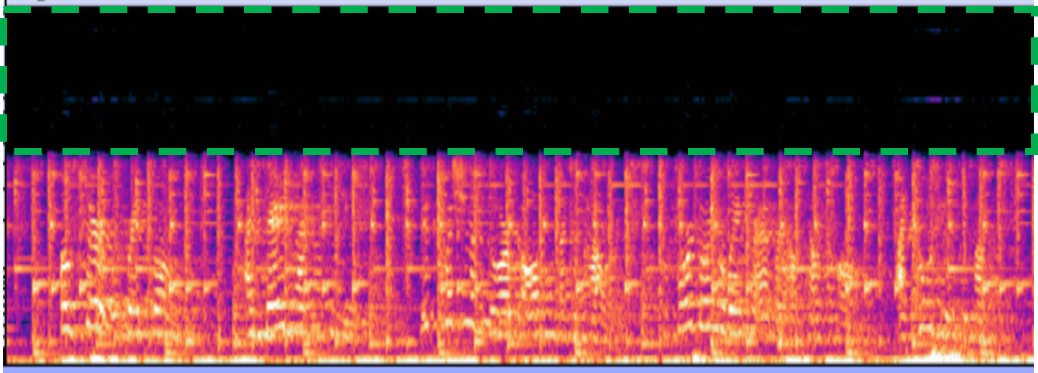}}
 \hfill
  \subfloat[Clean]{\includegraphics[width=0.45\linewidth, height=4cm]{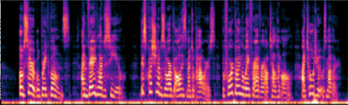}}
 \newline
  \subfloat[Regression model output]{\includegraphics[width=0.45\linewidth ,height=4cm]{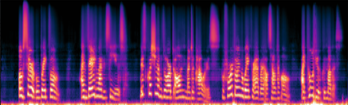}}
 \hfill
  \subfloat[GAN correction output (residual)]{\includegraphics[width=0.45\linewidth, height=4cm]{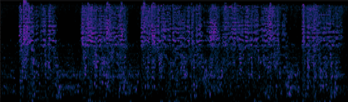}}
  \caption{Example illustrating that GANs can focus on correcting over-smoothed regions while leaving other parts unchanged. The noisy speech is \textbf{bandwidth-limited} in the green box, corresponding to a less informative region.}
  \label{fig: GAN correction1}
\end{figure*}

\begin{figure*}[t!]
 \subfloat[Noisy]{\includegraphics[width=0.45\linewidth, height=4cm]{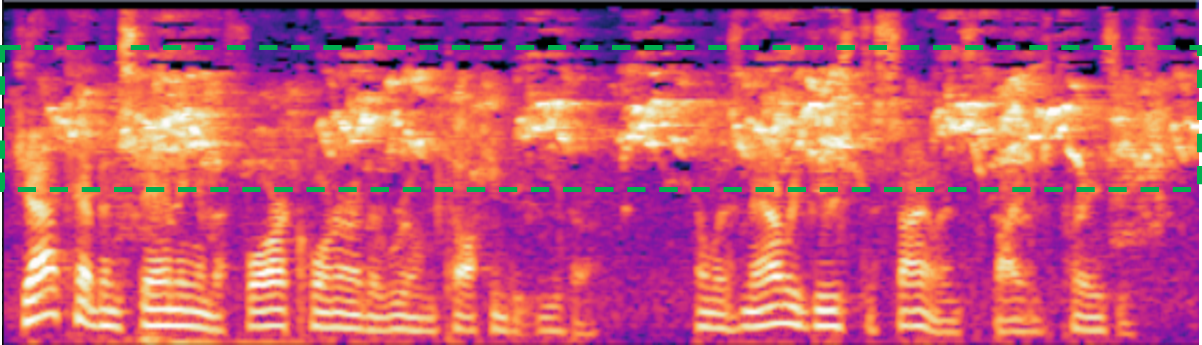}}
 \hfill
  \subfloat[Clean]{\includegraphics[width=0.45\linewidth, height=4cm]{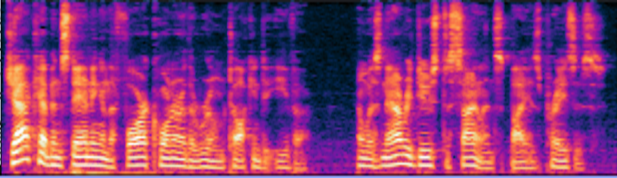}}
 \newline
  \subfloat[Regression model output]{\includegraphics[width=0.45\linewidth ,height=4cm]{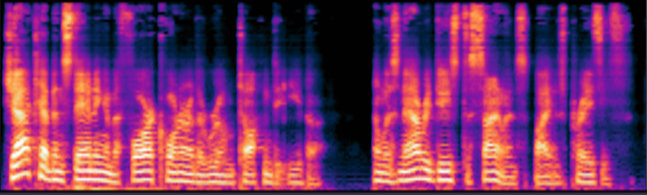}}
 \hfill
  \subfloat[GAN correction output (residual)]{\includegraphics[width=0.45\linewidth, height=4cm]{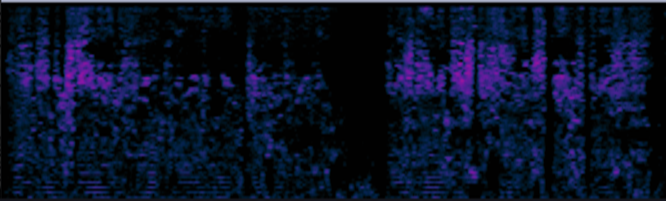}}
  \caption{Example illustrating that GANs can focus on correcting over-smoothed regions while leaving other parts unchanged. The noisy speech contains \textbf{strong noise} in the green box, corresponding to a less informative region.}
  \label{fig: GAN correction2}
\end{figure*}

\begin{figure*}[t]
 \subfloat[CommonVoice]{\includegraphics[width=0.43\linewidth, height=4.2cm]{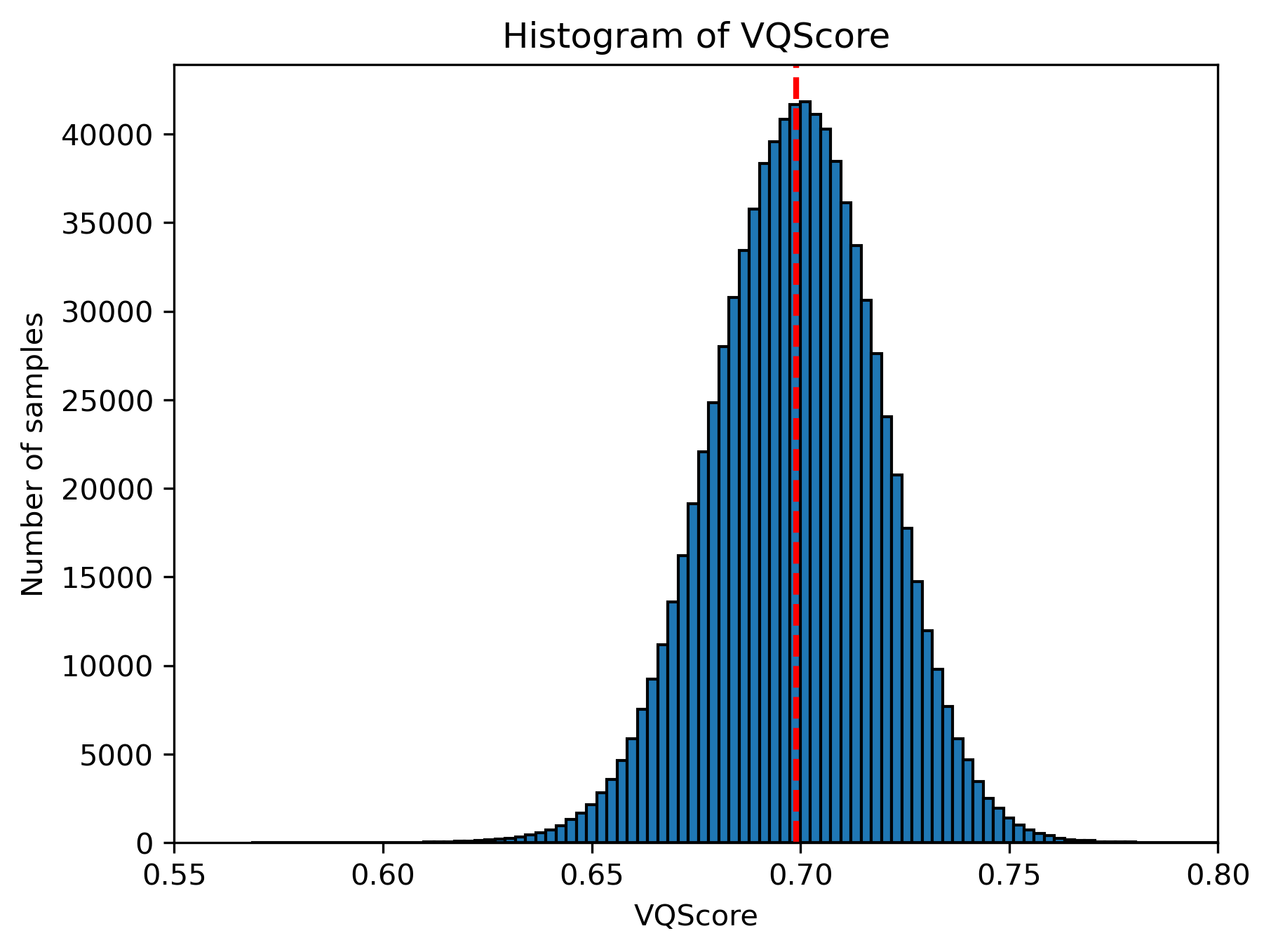}}
 \hfill
  \subfloat[DNS5]{\includegraphics[width=0.43\linewidth, height=4.2cm]{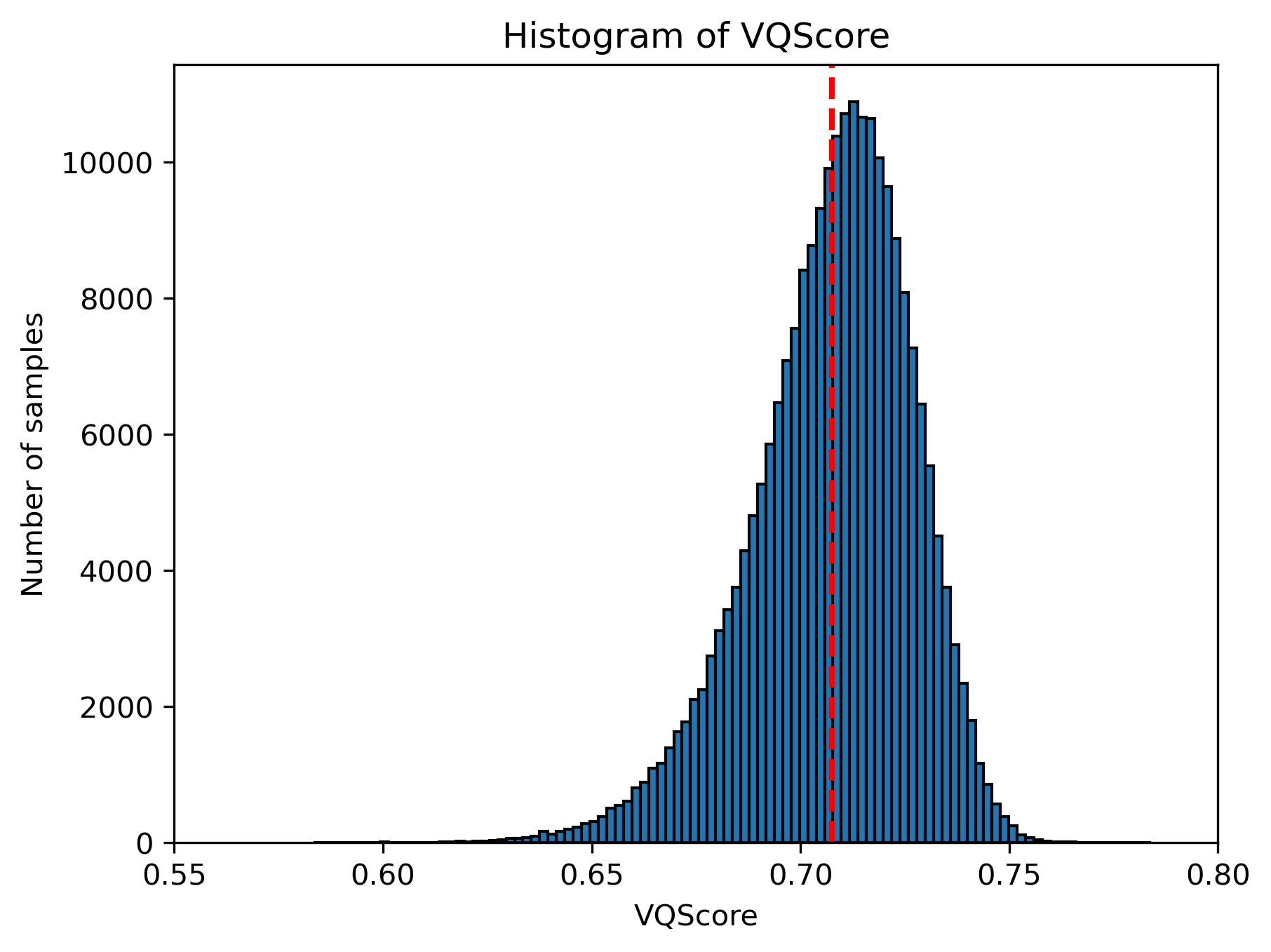}}
 \newline
  \subfloat[MLS]{\includegraphics[width=0.43\linewidth ,height=4.2cm]{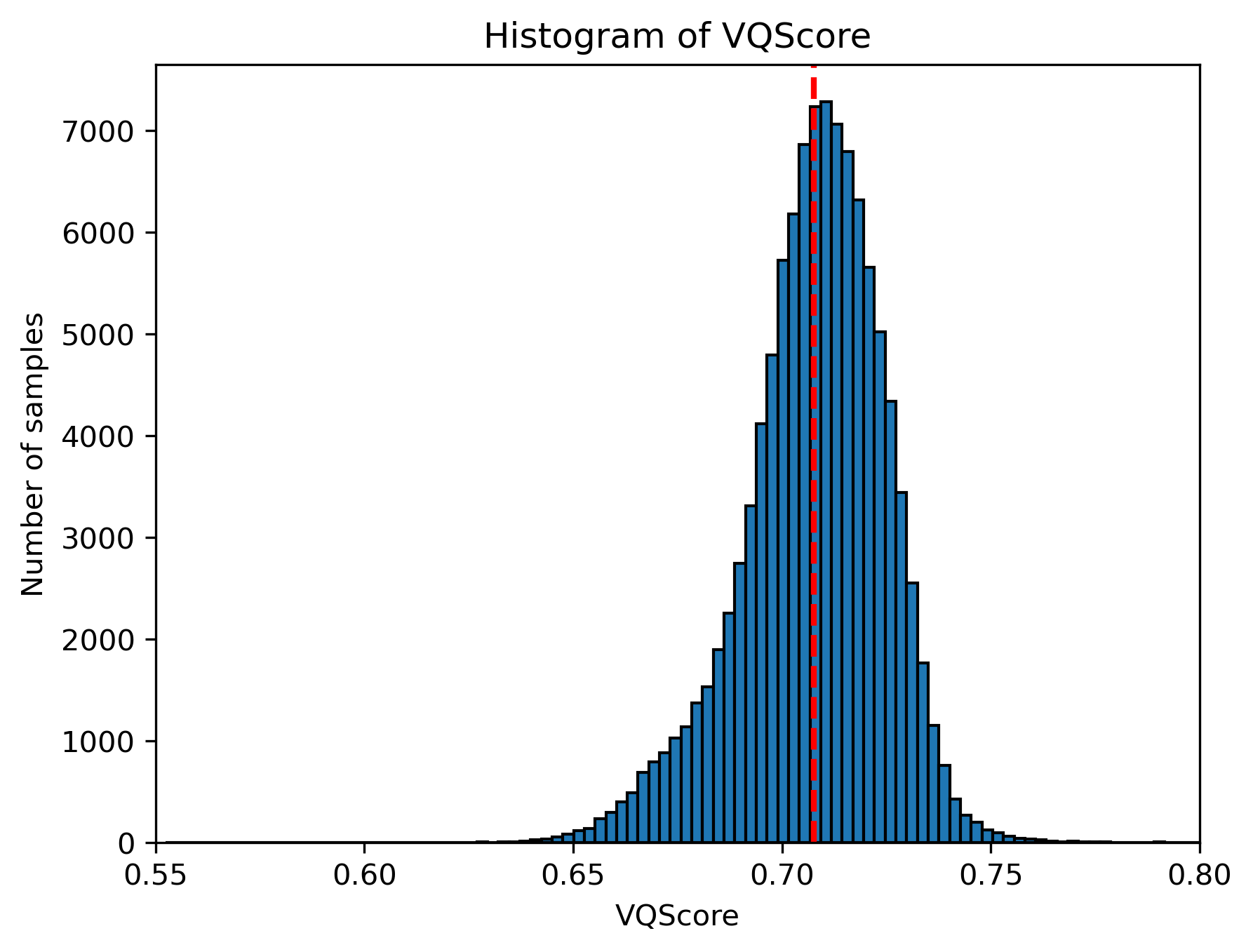}}
 \hfill
  \subfloat[LibriTTS]{\includegraphics[width=0.43\linewidth, height=4.2cm]{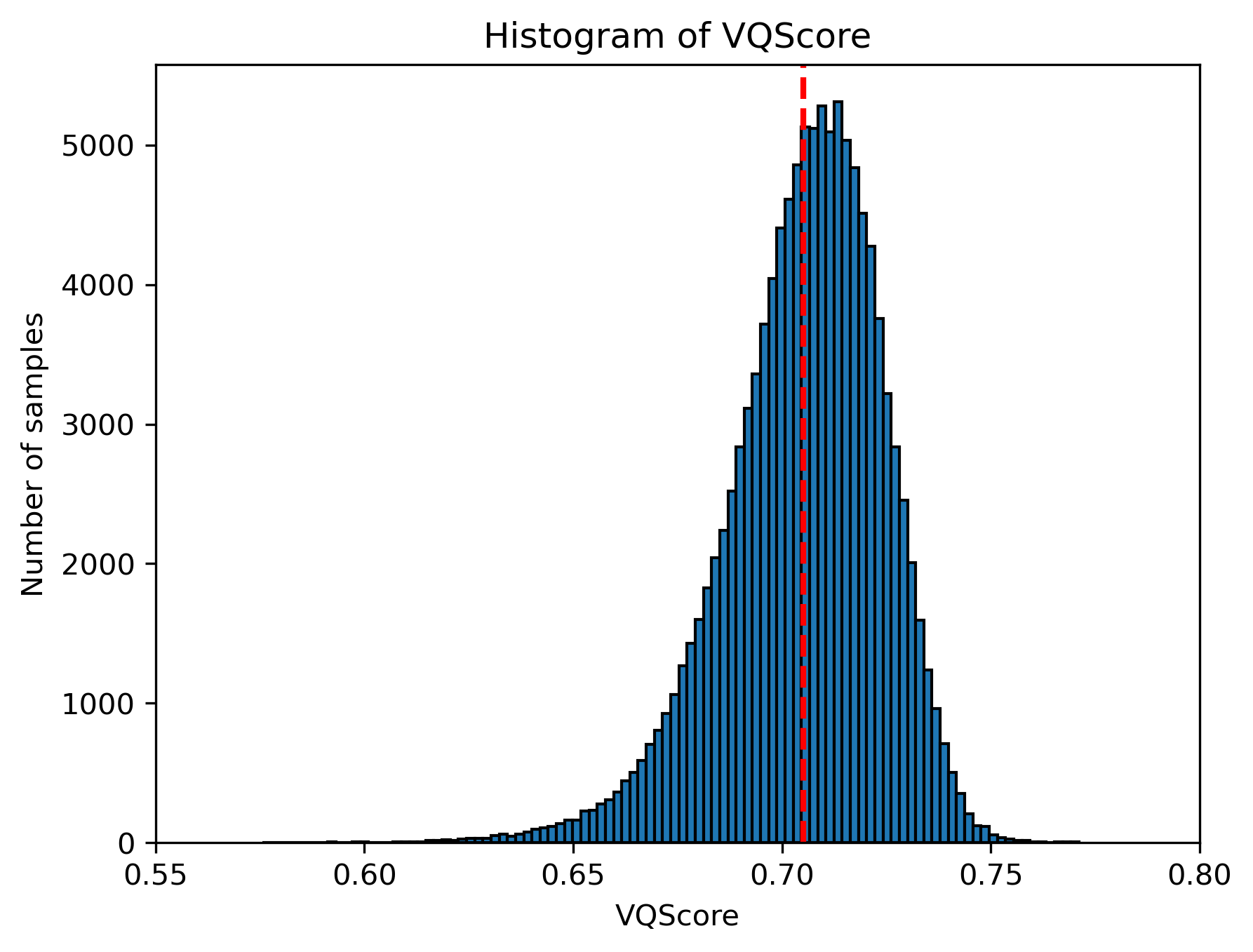}}

\subfloat[VCTK]{\includegraphics[width=0.43\linewidth ,height=4.2cm]{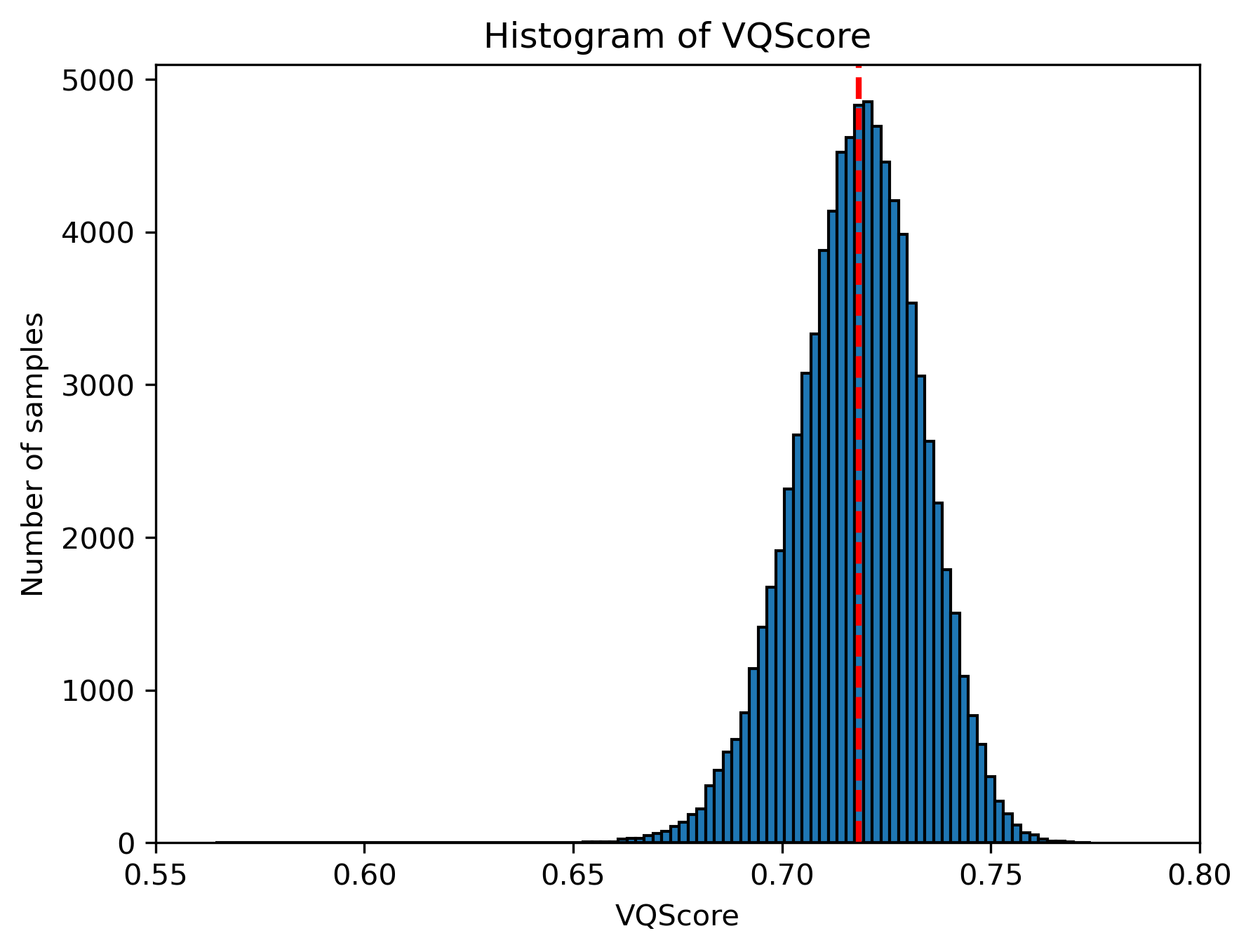}}
 \hfill
  \subfloat[WSJ]{\includegraphics[width=0.43\linewidth, height=4.2cm]{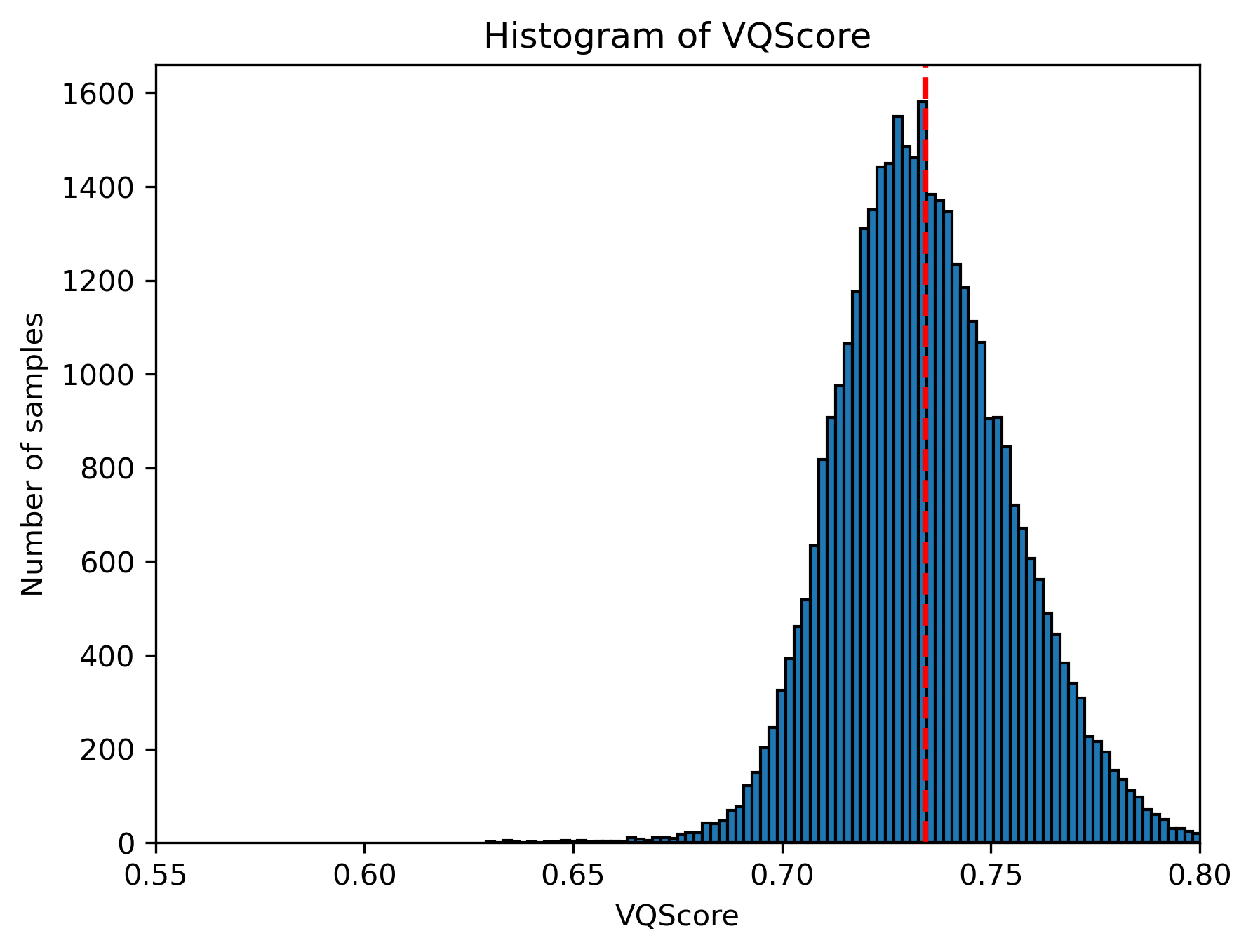}}
  \newline
  \subfloat[EARS]{\includegraphics[width=0.43\linewidth ,height=4.2cm]{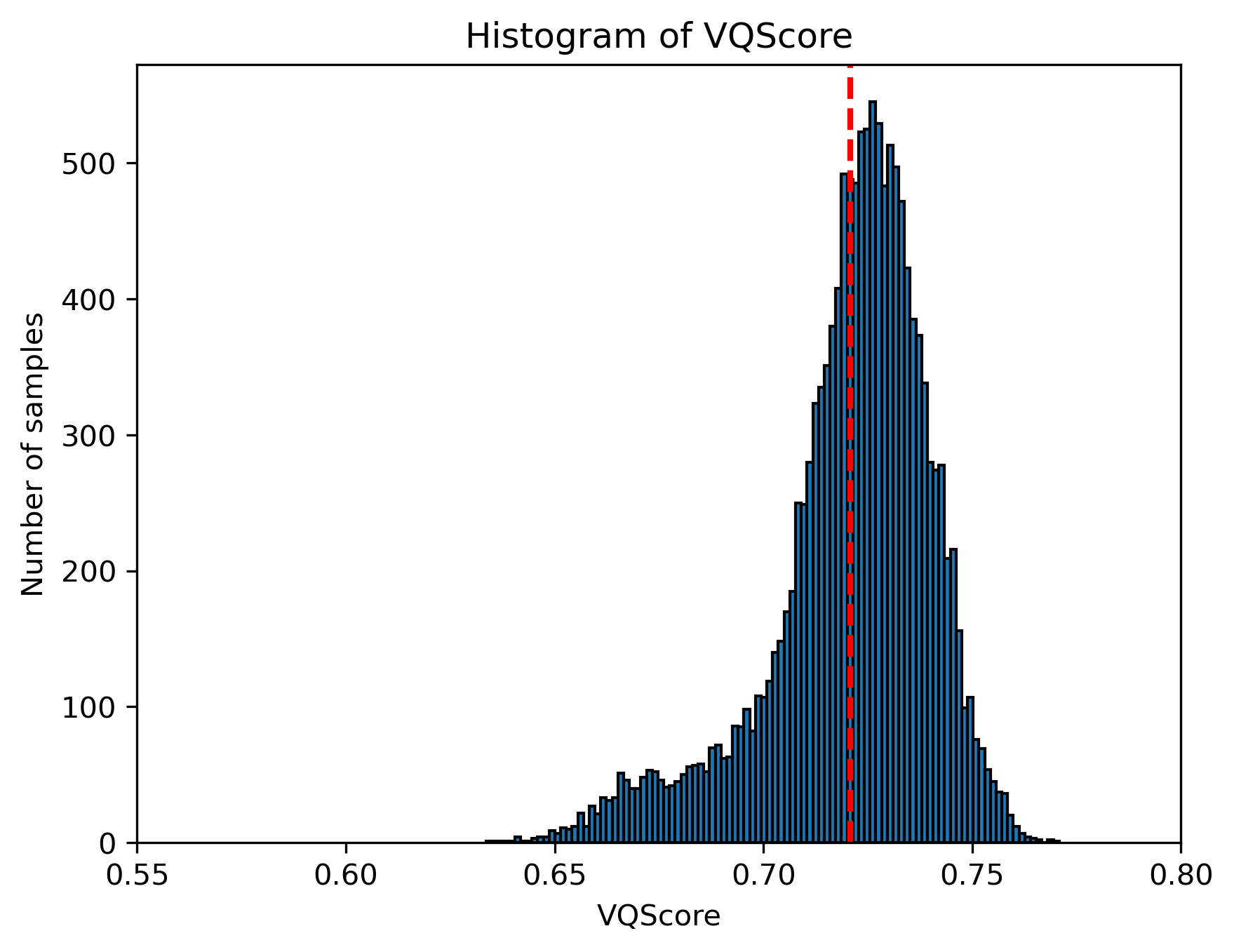}}
 \hfill
  \caption{Histogram of VQScore across different speech sources in the URGENT 2025 Challenge Track 1. The median of each data source is indicated by a dashed vertical line.}
  \label{fig: data_quality}
\end{figure*}

\begin{figure*}[t]
 \subfloat[Anechoic 1]{\includegraphics[width=0.45\linewidth, height=4cm]{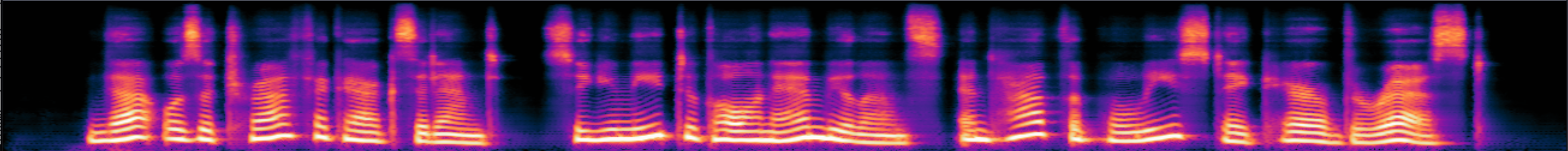}}
 \hfill
  \subfloat[Early reflected 1]{\includegraphics[width=0.45\linewidth, height=4cm]{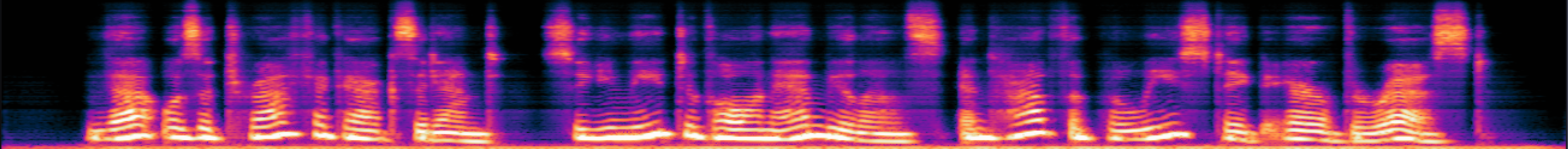}}
 \newline
  \subfloat[Anechoic 2]{\includegraphics[width=0.45\linewidth ,height=4cm]{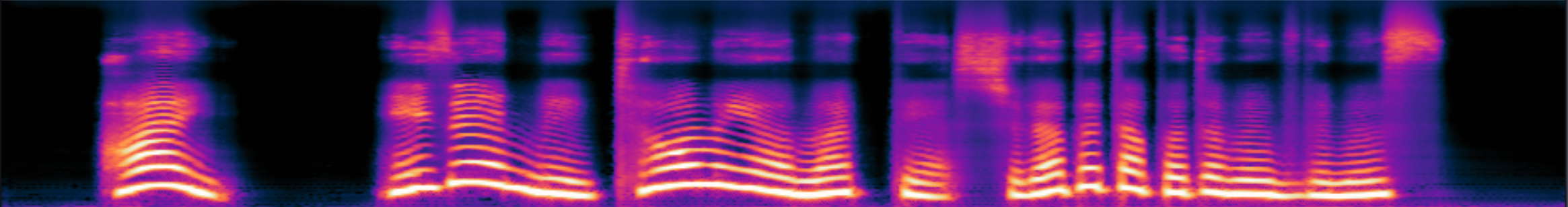}}
 \hfill
  \subfloat[Early reflected 2]{\includegraphics[width=0.45\linewidth, height=4cm]{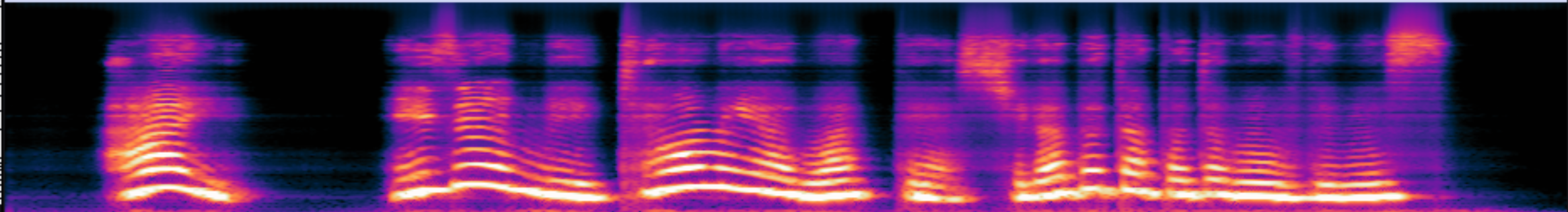}}
  \caption{Enhanced spectrogram comparison between using time-shifted anechoic clean speech and early-reflected speech as learning targets. (a) and (b) correspond to the same noisy input, and (c) and (d) correspond to another noisy input. Both samples are drawn from the blind-test set.}
  \label{fig: shifted anechoic clean speech VS early-reflected}
\end{figure*}

\begin{figure*}[t]
 \subfloat[Magnitude loss]{\includegraphics[width=0.48\linewidth, height=5.2cm]{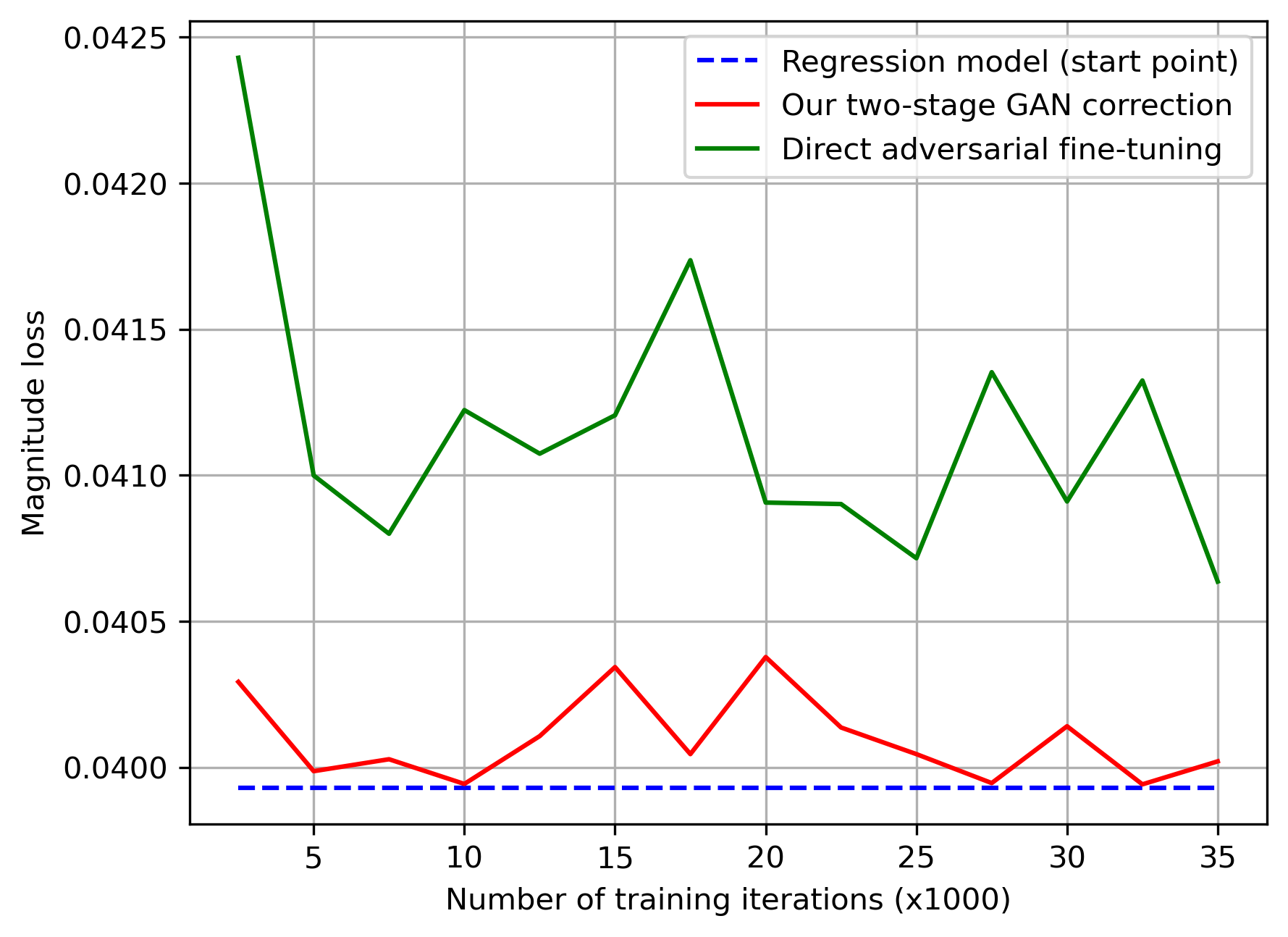}}
 \hfill
  \subfloat[Phase loss]{\includegraphics[width=0.48\linewidth, height=5.2cm]{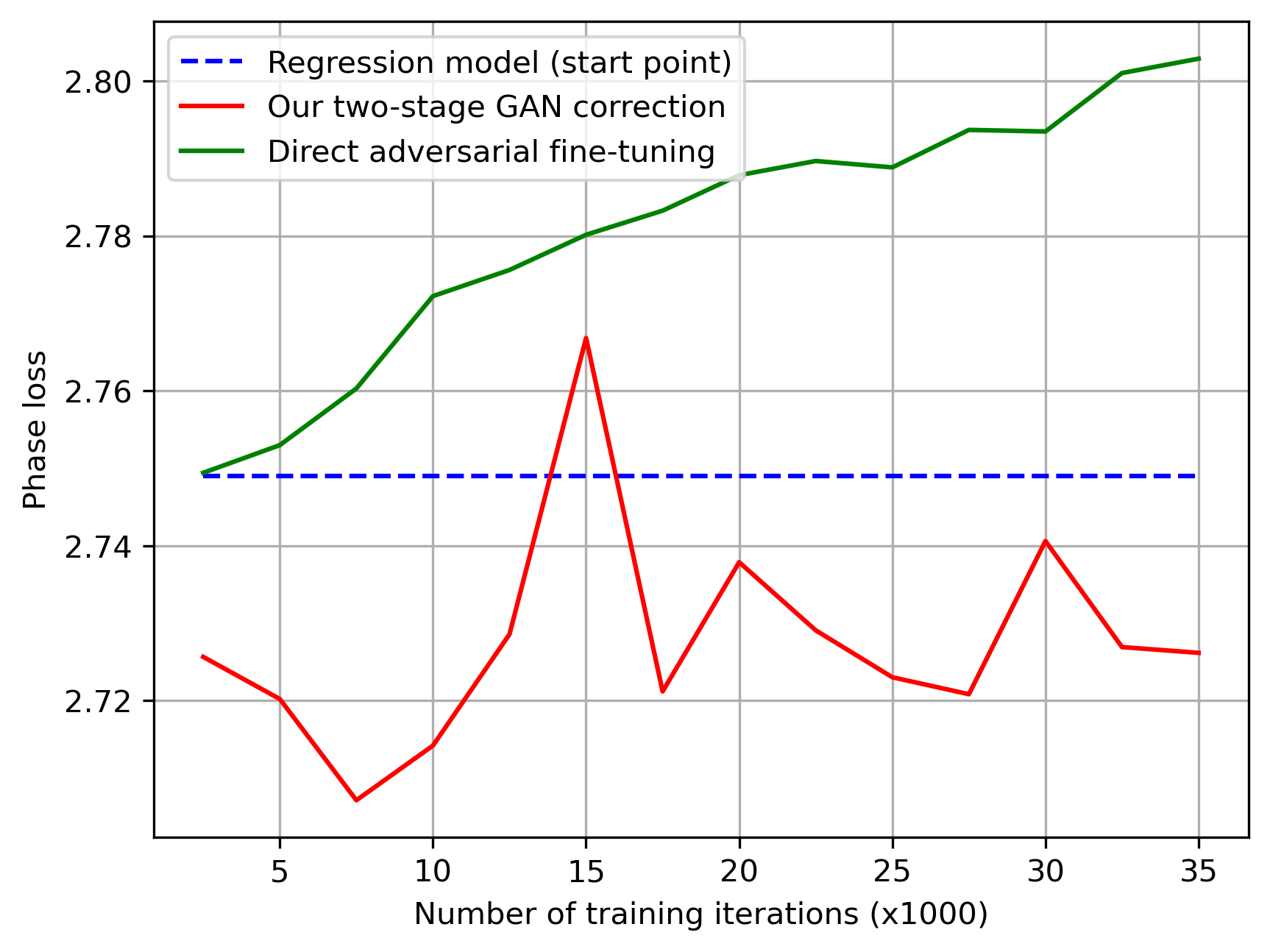}}
 \newline
  \subfloat[Time loss]{\includegraphics[width=0.48\linewidth ,height=5.2cm]{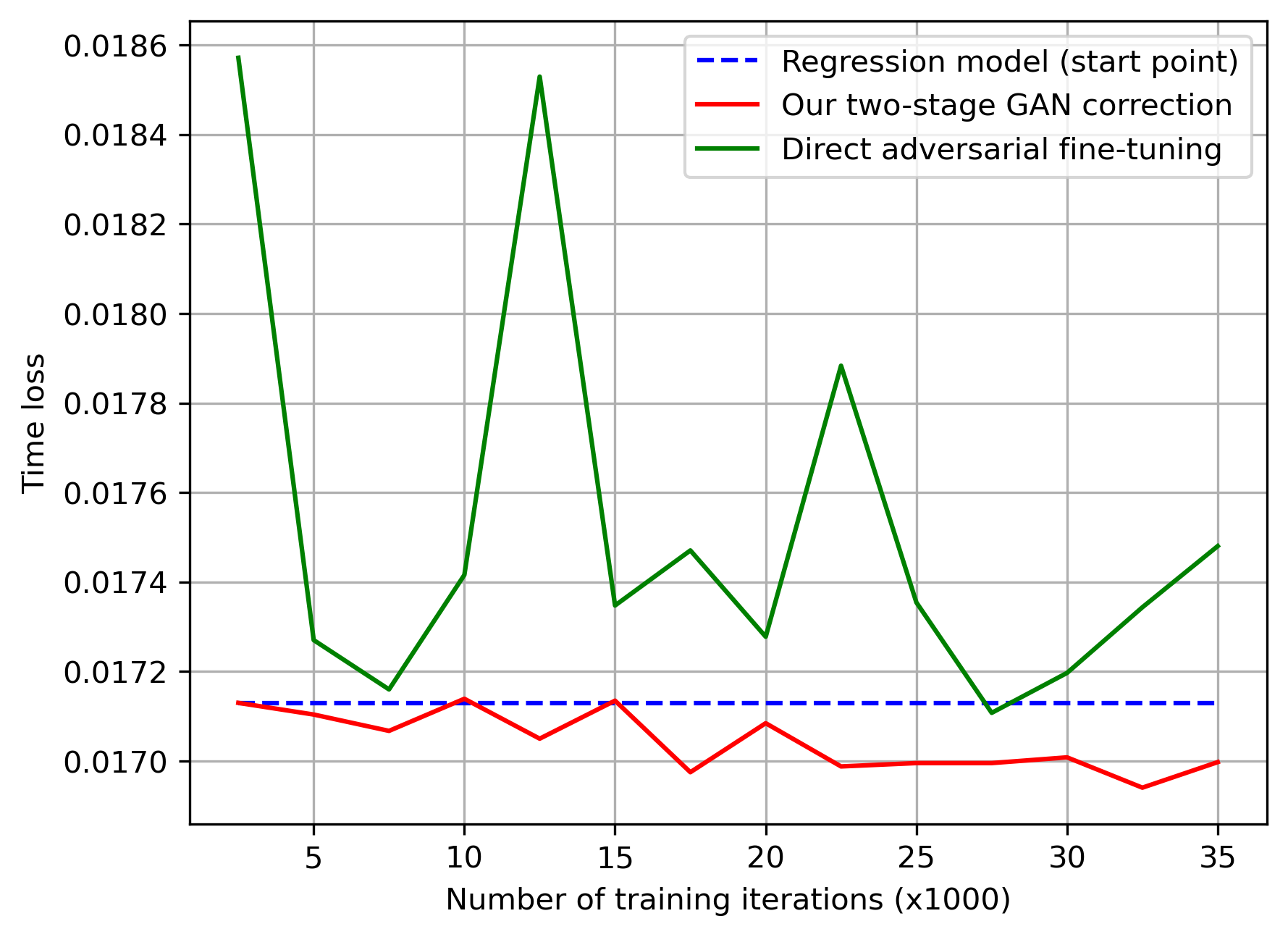}}
 \hfill
  \subfloat[PESQ Score]{\includegraphics[width=0.48\linewidth ,height=5.2cm]{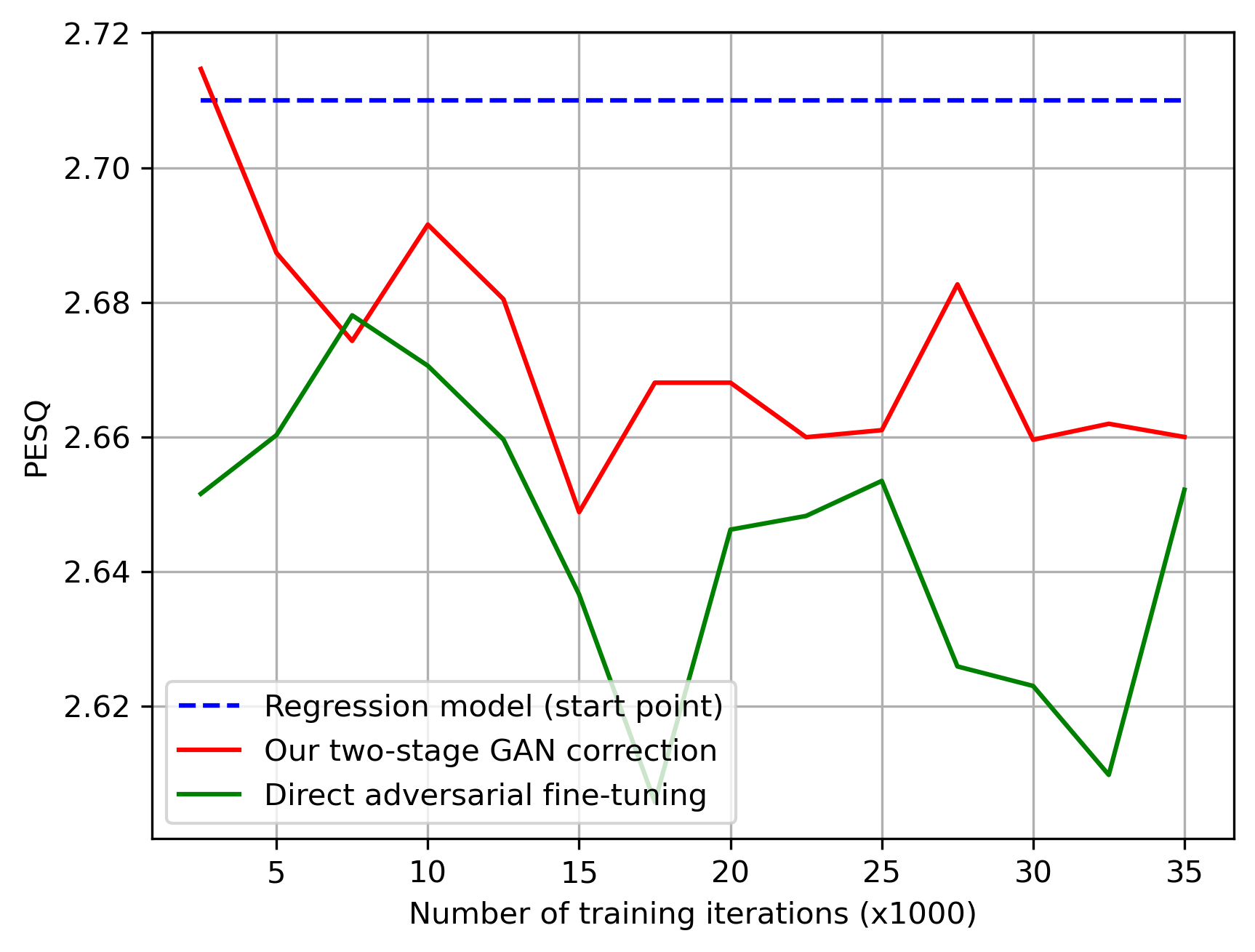}}
\caption{Learning curves comparison on validation-set between pre-training with a regression loss followed by adversarial fine-tuning and our two-stage GAN correction. (a) Magnitude loss, (b) Phase loss, (c) Time loss, and (d) PESQ score.}
\label{fig: direct  fine-tuning vs two-stage GAN}
\end{figure*}

\begin{figure*}[t]
 \subfloat[Original]{\includegraphics[width=0.45\linewidth, height=4cm]{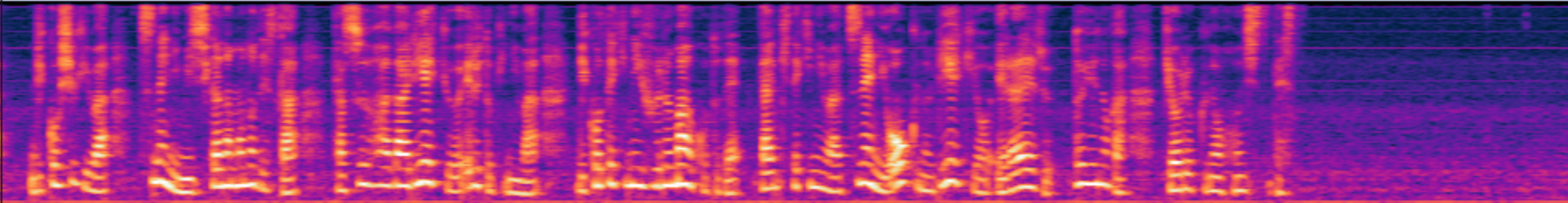}}
 \hfill
  \subfloat[FLEURS-R]{\includegraphics[width=0.45\linewidth, height=4cm]{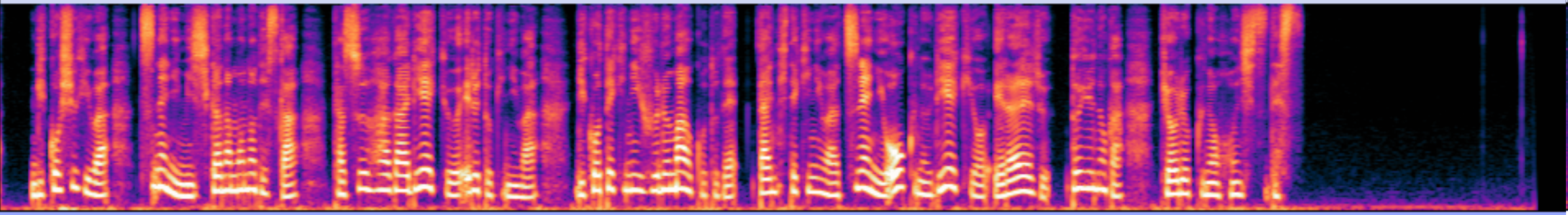}}
 \newline
  \subfloat[Proposed]{\includegraphics[width=0.45\linewidth ,height=4cm]{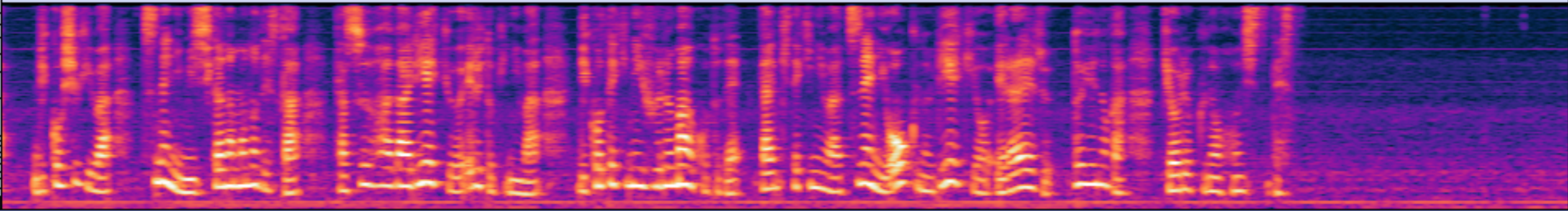}}
 \hfill
  \subfloat[Proposed (EARS)]{\includegraphics[width=0.45\linewidth, height=4cm]{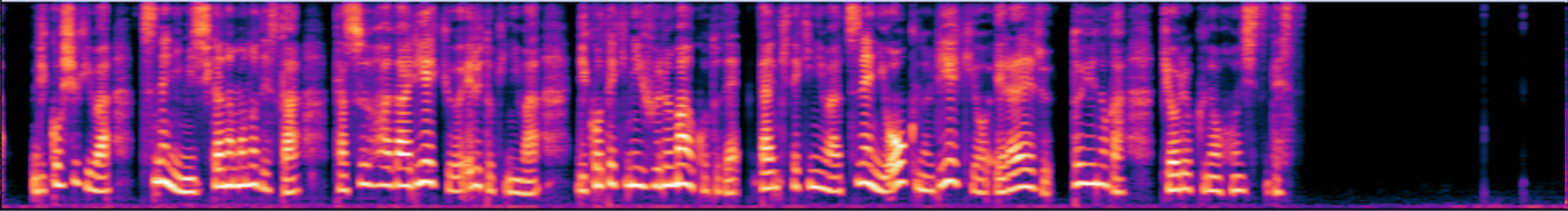}}
  \caption{Spectrogram comparison of a Japanese utterance (9997427445140542468.wav) from the FLEURS dataset. The original speech contains some very low-level stationary noise, which is commonly found in non-curated 'clean' training data.
  }
  \label{FLEURS}
\end{figure*}

\begin{table*}[t]
\centering
\caption{The overall ranking leaderboard for the non-blind test set of the URGENT 2025 Challenge (considering only the early reflected learning target for consistency).}
\resizebox{\textwidth}{!}{
\begin{tabular}
{lcccccccccccccc}
\toprule
Team & DNSMOS & NISQA & UTMOS & PESQ & ESTOI & SDR & MCD & LSD & SBERT & LPS & SpkSim & CAcc & Overall ranking \\
\midrule
Bobbsun & 3.01 (8) & 3.41 (6) & 2.40 (3) & 2.95 (1) & 0.86 (1) & 14.33 (1) & 3.01 (4) & 2.83 (5) & 0.91 (1) & 0.86 (1) & 0.85 (1) & 88.92 (1) & 2.516 \\

\textbf{Our Early reflected + GAN} & 3.04 (5) & 3.53 (3) & 2.30 (6) & 2.78 (4) & 0.84 (4) & 12.25 (5) & 2.97 (3) & 2.75 (4) & 0.90 (2) & 0.84 (3) & 0.85 (1) & 88.13 (2) & 3.166 \\

rc & 3.01 (8) & 3.21 (9) & 2.30 (6) & 2.79 (3) & 0.85 (2) & 13.11 (2) & 2.93 (2) & 2.94 (8) & 0.90 (2) & 0.85 (2) & 0.84 (3) & 88.05 (3) & 4.016 \\

\textbf{Our Early reflected} & 3.06 (4) & 3.23 (8) & 2.26 (8) & 2.81 (2) & 0.85 (2) & 12.28 (4) & 2.87 (1) & 2.66 (1) & 0.90 (2) & 0.84 (3) & 0.82 (5) & 87.62 (5) & 4.041 \\

Xiaobin & 3.00 (10) & 3.45 (4) & 2.31 (5) & 2.74 (5) & 0.84 (4) & 13.06 (3) & 3.30 (6) & 3.08 (11) & 0.89 (5) & 0.84 (3) & 0.83 (4) & 87.94 (4) & 5.033 \\

subatomicseer & 3.02 (6) & 3.28 (7) & 2.34 (4) & 2.63 (7) & 0.82 (6) & 12.18 (6) & 3.90 (12) & 3.06 (10) & 0.88 (7) & 0.82 (6) & 0.82 (5) & 86.15 (7) & 6.591 \\

poisonous & 3.02 (6) & 3.42 (5) & 2.26 (8) & 2.72 (6) & 0.82 (6) & 11.93 (8) & 3.36 (9) & 2.69 (2) & 0.89 (5) & 0.81 (8) & 0.80 (8) & 85.65 (8) & 6.758 \\

byti.shsy & 2.96 (12) & 3.15 (10) & 2.18 (10) & 2.44 (9) & 0.82 (6) & 12.09 (7) & 3.28 (5) & 3.27 (12) & 0.88 (7) & 0.82 (6) & 0.82 (5) & 86.46 (6) & 7.616 \\

Lam-Fung & 2.97 (11) & 2.95 (12) & 2.11 (15) & 2.43 (10) & 0.80 (9) & 11.37 (9) & 3.32 (7) & 2.84 (6) & 0.86 (10) & 0.79 (9) & 0.80 (8) & 84.93 (10) & 9.842 \\

urgent & 2.94 (13) & 2.89 (13) & 2.11 (15) & 2.43 (10) & 0.80 (9) & 11.29 (10) & 3.32 (7) & 2.85 (7) & 0.86 (10) & 0.79 (9) & 0.80 (8) & 84.96 (9) & 10.066 \\

cobalamin & 3.11 (3) & 2.70 (17) & 2.15 (11) & 2.22 (14) & 0.71 (16) & 6.22 (16) & 3.44 (10) & 2.72 (3) & 0.87 (9) & 0.79 (9) & 0.78 (12) & 83.73 (11) & 10.658 \\

alindborg & 3.28 (1) & 3.96 (2) & 2.49 (2) & 1.99 (15) & 0.76 (13) & 7.49 (15) & 4.51 (15) & 3.73 (14) & 0.84 (14) & 0.77 (14) & 0.77 (13) & 81.70 (14) & 10.891 \\

SQuad & 2.91 (16) & 2.89 (13) & 2.06 (17) & 2.35 (12) & 0.80 (9) & 10.93 (11) & 3.57 (11) & 3.03 (9) & 0.86 (10) & 0.79 (9) & 0.79 (11) & 83.71 (12) & 11.683 \\

dy & 2.93 (15) & 2.97 (11) & 2.14 (13) & 2.56 (8) & 0.78 (12) & 9.58 (12) & 4.40 (13) & 3.28 (13) & 0.85 (13) & 0.78 (13) & 0.77 (13) & 83.20 (13) & 12.650 \\

wataru9871 & 3.18 (2) & 4.01 (1) & 2.78 (1) & 1.36 (19) & 0.56 (19) & -13.88 (19) & 11.25 (19) & 7.98 (19) & 0.82 (17) & 0.73 (17) & 0.51 (19) & 79.70 (18) & 13.958 \\

IASP\_Q & 2.94 (13) & 2.77 (16) & 2.14 (13) & 2.25 (13) & 0.76 (13) & 6.00 (17) & 5.21 (16) & 4.31 (15) & 0.83 (15) & 0.74 (15) & 0.69 (16) & 80.03 (17) & 15.075 \\

hanhw96 & 2.63 (18) & 2.42 (18) & 1.87 (18) & 1.91 (17) & 0.72 (15) & 8.28 (13) & 4.41 (14) & 4.89 (16) & 0.83 (15) & 0.74 (15) & 0.70 (15) & 81.66 (15) & 15.750 \\

SEES & 2.88 (17) & 2.80 (15) & 2.15 (11) & 1.99 (15) & 0.68 (17) & 8.07 (14) & 5.78 (17) & 6.59 (18) & 0.79 (18) & 0.66 (18) & 0.54 (18) & 71.74 (19) & 16.758 \\

noisy & 1.84 (19) & 1.69 (19) & 1.56 (19) & 1.37 (18) & 0.61 (18) & 2.53 (18) & 7.92 (18) & 5.51 (17) & 0.75 (19) & 0.62 (19) & 0.63 (17) & 81.29 (16) & 18.075 \\
\bottomrule
\end{tabular}
}
\label{tab:challenge_results}
\end{table*}

\begin{table*}[t]
    \caption{Zero-shot TTS evaluation after training data cleaning using our USE model on unseen languages. We report the 95\% confidence intervals based on standard errors calculated from 10 independent runs per dataset.}
    \centering
    \small
    \setlength{\tabcolsep}{4pt}
    \begin{tabular}{cccrccc}
        \toprule
        \textbf{Language} & \textbf{Context Audio} & \textbf{Train Audio} & \textbf{CER (\%)} & \textbf{WER (\%)} & \textbf{SpkSim} & \textbf{FCD} \\
        \midrule
        Dutch & original & original & 14.28 $\pm$ 0.98 &  19.60 $\pm$ 0.76 & 0.6064 $\pm$ 0.0080 & 0.2444 $\pm$ 0.0155 \\
              & enhanced & original & 12.93 $\pm$ 0.48 &  19.21 $\pm$ 0.63 & \textbf{0.6643} $\pm$ 0.0096 & 0.2282 $\pm$ 0.0135 \\
              & original & enhanced & 8.24 $\pm$ 0.55 &  14.31 $\pm$ 0.69 & 0.6359 $\pm$ 0.0050 & \textbf{0.1761} $\pm$ 0.0070 \\
              & enhanced & enhanced & \textbf{7.75} $\pm$ 0.83 & \textbf{13.66} $\pm$ 0.71 & 0.6603 $\pm$ 0.0047 & 0.1837 $\pm$ 0.0086 \\
        \midrule
        Italian & original & original & 11.13 $\pm$ 0.94 & 19.20 $\pm$ 0.94 & 0.6004 $\pm$ 0.0034 & 0.1846 $\pm$ 0.0042 \\
                & enhanced & original & 12.18 $\pm$ 0.52 & 20.41 $\pm$ 0.75 & \textbf{0.6135} $\pm$ 0.0047 & \textbf{0.1321} $\pm$ 0.0046 \\
                & original & enhanced & 10.56 $\pm$ 0.45 & 18.91 $\pm$ 0.43 & 0.5869 $\pm$ 0.0054 & 0.2045 $\pm$ 0.0041 \\
                & enhanced & enhanced & \textbf{8.30} $\pm$ 0.52 & \textbf{15.98} $\pm$ 0.53 & 0.6006 $\pm$ 0.0032 & 0.1373 $\pm$ 0.0021 \\
        \bottomrule
    \end{tabular}
    \label{tab:additional_tts}
\end{table*}

\clearpage
\section*{NeurIPS Paper Checklist}

\begin{enumerate}

\item {\bf Claims}
    \item[] Question: Do the main claims made in the abstract and introduction accurately reflect the paper's contributions and scope?
    \item[] Answer: \answerYes{} 
    \item[] Justification: We have clearly described contributions and scope in the abstract and introduction.

\item {\bf Limitations}
    \item[] Question: Does the paper discuss the limitations of the work performed by the authors?
    \item[] Answer: \answerYes{} 
    \item[] Justification: We provide the limitations in the Appendix (see the “Limitations” section).

\item {\bf Theory assumptions and proofs}
    \item[] Question: For each theoretical result, does the paper provide the full set of assumptions and a complete (and correct) proof?
    \item[] Answer: \answerYes{} 
    \item[] Justification: 
    Formal proofs are provided in Appendix A1.

    \item {\bf Experimental result reproducibility}
    \item[] Question: Does the paper fully disclose all the information needed to reproduce the main experimental results of the paper to the extent that it affects the main claims and/or conclusions of the paper (regardless of whether the code and data are provided or not)?
    \item[] Answer: \answerYes{} 
    \item[] Justification: 
    We have described the model architecture and experiment setting carefully.

\item {\bf Open access to data and code}
    \item[] Question: Does the paper provide open access to the data and code, with sufficient instructions to faithfully reproduce the main experimental results, as described in supplemental material?
    \item[] Answer: \answerNo{} 
    \item[] Justification: We will release the code and model after the paper has been accepted.

\item {\bf Experimental setting/details}
    \item[] Question: Does the paper specify all the training and test details (e.g., data splits, hyperparameters, how they were chosen, type of optimizer) necessary to understand the results?
    \item[] Answer: \answerYes{} 
    \item[] Justification: The training and test details are provided in the Experiments section.

\item {\bf Experiment statistical significance}
    \item[] Question: Does the paper report error bars suitably and correctly defined or other appropriate information about the statistical significance of the experiments?
    \item[] Answer: \answerYes{} 
    \item[] Justification: We report  95\% confidence
    intervals for the TTS results and the subjective listening tests. Due to the high computational cost, we do not report statistical significance for the general USE results.

\item {\bf Experiments compute resources}
    \item[] Question: For each experiment, does the paper provide sufficient information on the computer resources (type of compute workers, memory, time of execution) needed to reproduce the experiments?
    \item[] Answer: \answerYes{} 
    \item[] Justification: We provide our computing requirements in the experiment section.
    
\item {\bf Code of ethics}
    \item[] Question: Does the research conducted in the paper conform, in every respect, with the NeurIPS Code of Ethics \url{https://neurips.cc/public/EthicsGuidelines}?
    \item[] Answer: \answerYes{} 
    \item[] Justification: We follow NeurIPS Code of Ethics to keep anonymity, etc.

\item {\bf Broader impacts}
    \item[] Question: Does the paper discuss both potential positive societal impacts and negative societal impacts of the work performed?
    \item[] Answer: \answerNA{} 
    \item[] Justification: For USE, the primary application is to improve communication with limited or no societal impact.
    
\item {\bf Safeguards}
    \item[] Question: Does the paper describe safeguards that have been put in place for responsible release of data or models that have a high risk for misuse (e.g., pre-trained language models, image generators, or scraped datasets)?
    \item[] Answer: \answerNA{} 
    \item[] Justification: We don't think our model needs safeguards to prevent misuse.

\item {\bf Licenses for existing assets}
    \item[] Question: Are the creators or original owners of assets (e.g., code, data, models), used in the paper, properly credited and are the license and terms of use explicitly mentioned and properly respected?
    \item[] Answer: \answerYes{} 
    \item[] Justification: The related information is explicitly mentioned and properly cited.

\item {\bf New assets}
    \item[] Question: Are new assets introduced in the paper well documented and is the documentation provided alongside the assets?
    \item[] Answer: \answerYes{} 
    \item[] Justification: All the audio demos provided on the demo page are properly cited.

\item {\bf Crowdsourcing and research with human subjects}
    \item[] Question: For crowdsourcing experiments and research with human subjects, does the paper include the full text of instructions given to participants and screenshots, if applicable, as well as details about compensation (if any)? 
    \item[] Answer: \answerYes{} 
    \item[] Justification: 
    This information is provided in the Appendix related to the subjective test.

\item {\bf Institutional review board (IRB) approvals or equivalent for research with human subjects}
    \item[] Question: Does the paper describe potential risks incurred by study participants, whether such risks were disclosed to the subjects, and whether Institutional Review Board (IRB) approvals (or an equivalent approval/review based on the requirements of your country or institution) were obtained?
    \item[] Answer: \answerYes{} 
    \item[] Justification: It is not required to get an IRB for our subjective test.

\item {\bf Declaration of LLM usage}
    \item[] Question: Does the paper describe the usage of LLMs if it is an important, original, or non-standard component of the core methods in this research? Note that if the LLM is used only for writing, editing, or formatting purposes and does \emph{not} impact the core methodology, scientific rigor, or originality of the research, declaration is not required.
    \item[] Answer: \answerNA{} 
    \item[] Justification: The core method development in this research does not involve LLMs as any important components.

\end{enumerate}

\end{document}